\documentclass{article}

\usepackage{arxiv}

\usepackage[utf8]{inputenc} 
\usepackage[T1]{fontenc}    
\usepackage{hyperref}       
\usepackage{url}            
\usepackage{booktabs}       
\usepackage{amsfonts}       
\usepackage{amsmath}
\usepackage{caption}
\usepackage{nicefrac}       
\usepackage{microtype}      
\usepackage{cleveref}       
\usepackage{graphicx}
\usepackage{doi}
\usepackage{cite}			

\title{Macroscopic Structural Light Absorbers}

\author{{\hspace{1mm}Jan M. Kaster}\\
	Corporate Research and Technology\\
	Carl Zeiss AG\\
	Oberkochen, Germany\\
	\texttt{jan.kaster@zeiss.com} \\
}

\renewcommand{\shorttitle}{Macroscopic Structural Light Absorbers}

\hypersetup{
pdftitle={Macroscopic Structural Light Absorbers},
pdfsubject={physics.optics, physics.app-ph, physics.ins-det},
pdfauthor={Jan M. Kaster},
pdfkeywords={Light absorber, Stray light suppression, Heat transfer, Light weighting, Additive manufacturing},
}

\begin{document}
\maketitle

\begin{abstract}
The interaction of light with optical and mechanical systems is influenced by material properties, geometrical configurations, and surface topographies. Designing these systems necessitates a careful balance of conflicting requirements, such as minimising size and weight while simultaneously improving heat transfer and reducing stray light from illuminated peripheral mounting surfaces. Stray light is typically mitigated by apertures, coatings, and microscopic structures, alongside maintaining cleanliness. However, using apertures may not always be feasible, and effective optical absorber coatings or microscopic light absorbing structures can be costly and sensitive to environmental factors such as abrasion, radiation heating, or cleaning agents.

In a proof-of-concept investigation, we design and analyse macroscopic structural light absorbers realised as periodic minimal surface approximations and quasi-stochastic lattices. The term "macroscopic" refers to minimal structural dimensions of approximately 100 micrometres. By increasing the number of reflections before residual reflected light reaches a hemispherical receiver, we achieve reductions in received peak intensities by factors of less than 0.39 and average intensities by factors of less than 0.65, without altering the surface properties.

Macroscopic structural light absorbers support cost-effective and robust light-absorbing materials, such as black anodised aluminium or ABS polymers, while still achieving satisfactory stray light suppression. This approach is applicable to aerospace optical systems (such as telescopes and imaging spectrometers), as well as general scientific and industrial optical instruments and commercial products (including projectors and luminaires). The demonstrated structures can be sustainably manufactured through processes like laser powder bed fusion, stereo-lithography, or fused deposition modelling.
\end{abstract}

\keywords{Light absorber \and Stray light suppression \and Heat transfer \and Light weighting \and Additive manufacturing}

\section*{Main}
\label{sec:main}
Optical systems may encounter performance limitations due to residual stray light. \cite{harvey1977light}, \cite{harvey2019understanding} The primary objective of radiometric optimisation in such systems is to minimise irradiating surfaces from which stray light can propagate to a critical location following a single reflection. \cite{breault1977problems}, \cite{fest2013straylight}, \cite{clermont2024corot} For instance, it is crucial to prevent light from directly scattering onto a detector from a mounting surface. However, due to stringent space and weight constraints, the control of stray light through state-of-the-art design methodologies can be limited, particularly concerning mounting and housing geometries. \cite{clermont2024metop}, \cite{chen2024deep}
    
As a general approach, the surfaces of peripheral geometries are treated to enhance light absorption, and the number of reflections is maximised before stray light propagates further through the optical system. This can be achieved through the application of coatings \cite{dobrowolski1995filters}, \cite{cai2022absorber}, \cite{luhmann2020gold} nanostructures, \cite{mizuno2009carbon}, \cite{aydin2011plasmonic} apertures, \cite{he2024baffle} or saw-tooth-like geometric shapes \cite{stavroudis1994baffles}, among other techniques. Concepts aimed at increasing the number of reflections to enhance absorption, referred to as structural light absorbers, can be observed in nature, such as the microscopic and nanoscopic structures found on the feathers of certain bird species \cite{mccoy2018feathers} or the wings of some butterfly species. \cite{vukusic2004butterfly}, \cite{davis2020ultrablack}, \cite{zhao2011carbon}
    
For coatings, and particularly for microscopic structural light absorbers with minimal structural dimensions comparable to or less than the wavelength of the light to be absorbed, a challenging trade-off exists in balancing optical performance with other system design objectives, such as material compatibility (e.g., adhesion), costs (including supply chain limitations and complex fabrication processes), and environmental durability and reliability (encompassing cleanliness, abrasion resistance, radiation heating, and vibration).
    
The objective of the present proof-of-concept investigation is to evaluate whether a macroscopic structural light absorber, designed using bio-inspired foam-like minimal surface approximations \cite{mackay1985minimal} and sponge-like lattice structures, \cite{fernandes2021lattices} can reduce its reflected intensity based on its geometry. The hypothesis posits that it is feasible to design structures that modify the reflected intensity in specific ways. For example, this could involve increasing the number of reflections within the structure before the reflected light propagates further into the optical system or directing scatter in a backward direction - towards the luminaire, rather than forward scatter. The latter approach can be advantageous in optical systems characterised by unidirectional light propagation. \cite{stavroudis1994baffles}

Furthermore, we aim to develop macroscopic structural light absorbers that can be additively manufactured and that offer opportunities for enhancements such as weight reduction, \cite{prathyusha2022review} local structural stiffness reduction, \cite{daynes2025graded} engineered thermal expansion \cite{chen2018metamaterials} and improved heat exchange \cite{careri2023heat}, \cite{wang2024tpms} or insulation. \cite{an2022insulation}
    
For this study, we designed three different variants of macroscopic structural light absorbers and compared them to a planar sample with similar optical surface properties. A commercial software package was employed for the generation of the geometries \cite{ansys2025spaceclaim} and the simulation-based optical analysis. \cite{ansys2025speos} The chosen geometries are based on approximations of periodic minimal surfaces of Gyroid and Schwarz D types, \cite{schoen1970minimal}, \cite{hoffman2001minimal} as well as a boundary-conforming lattice consisting of cylindrical struts that follow the edges of two intertwined tetrahedra and an octahedron centred on these tetrahedra. This lattice is referred to as the "Double Pyramid and Face Diagonals" lattice in the commercial design software. The boundary-conforming lattice generation ensures that the structure adheres to the initial body shape, which can result in a quasi-stochastic pattern, even for a simple initial geometry such as a cylindrical disc.
    
\begin{figure}[htbp]
    \centering

    \begin{minipage}[b]{0.3\textwidth}
        \centering
        \includegraphics[height=3.5cm]{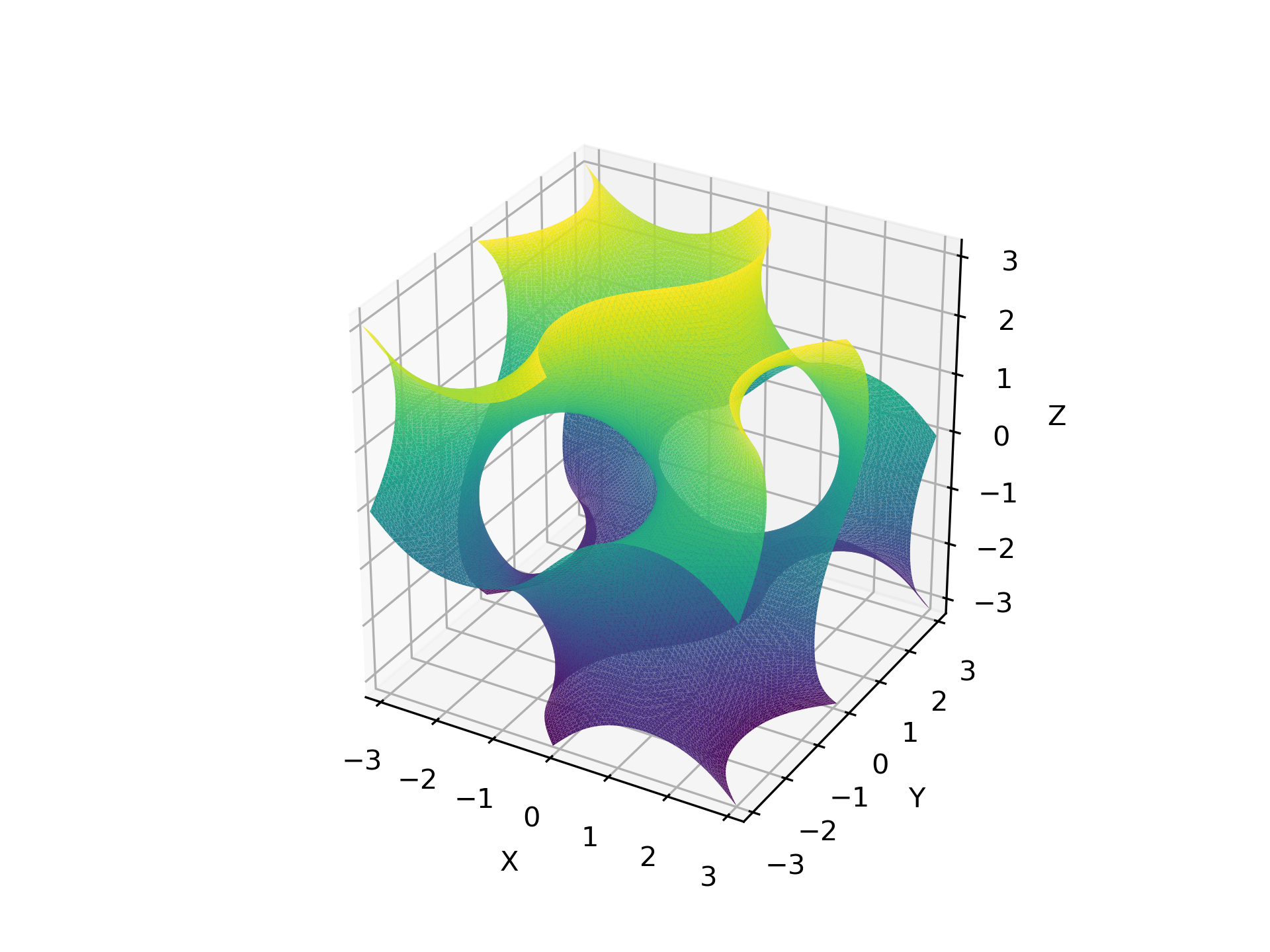}
    \end{minipage}
    \hfill
    \begin{minipage}[b]{0.3\textwidth}
        \centering
        \includegraphics[height=3.5cm]{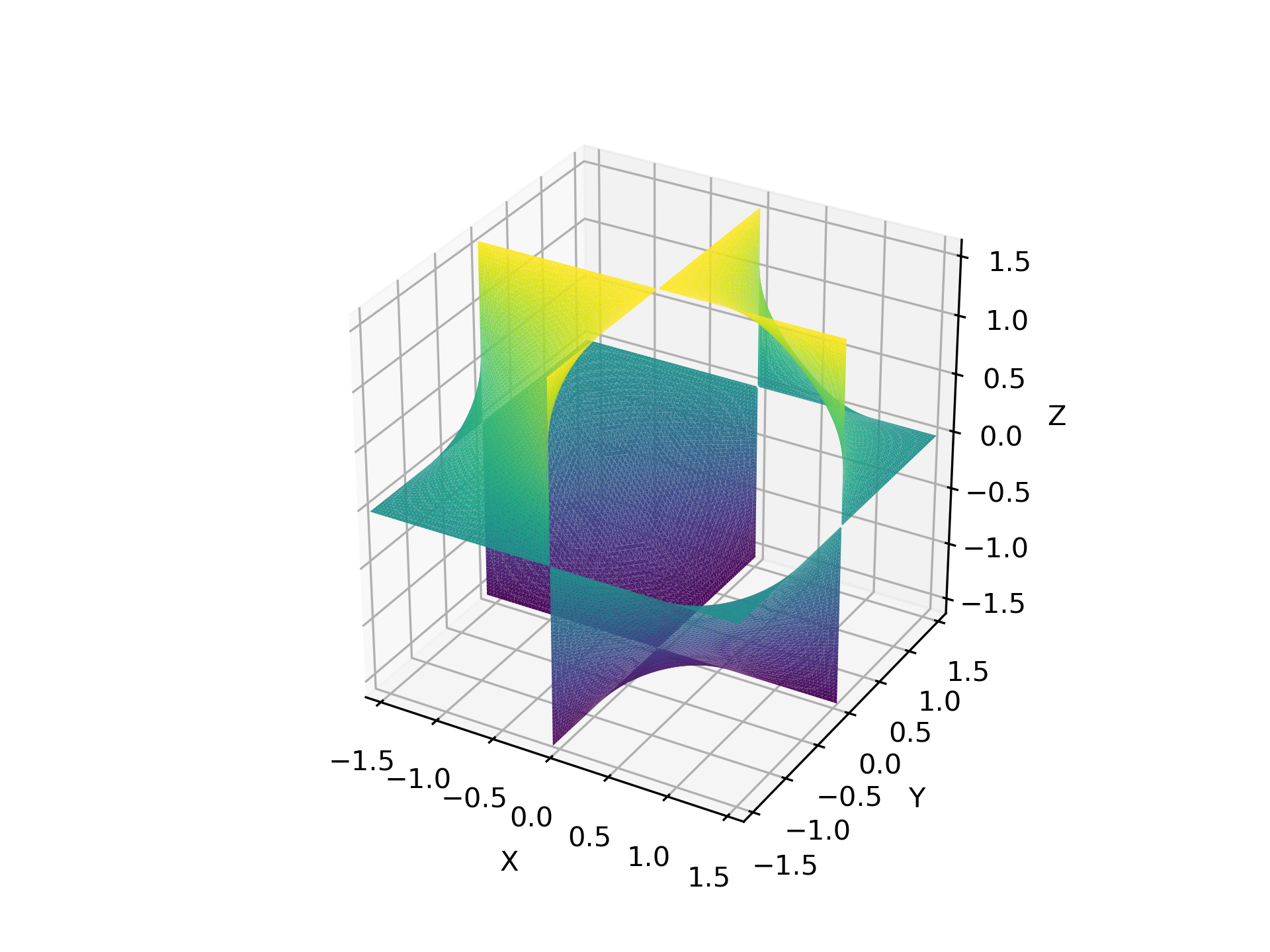}
    \end{minipage}
    \hfill
    \begin{minipage}[b]{0.3\textwidth}
        \centering
        \includegraphics[height=3cm]{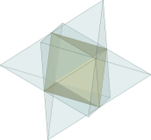}
    \end{minipage}

    \vspace{0.5cm} 

    \begin{minipage}[b]{0.3\textwidth}
        \centering
        \includegraphics[height=4cm]{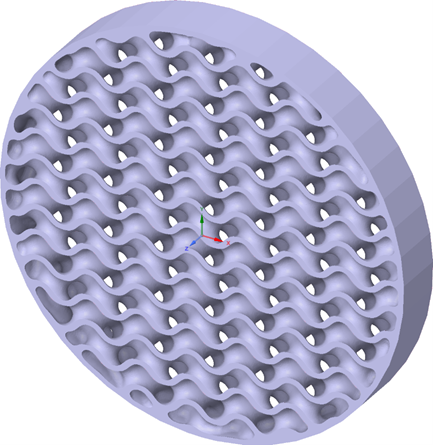}
    \end{minipage}
    \hfill
    \begin{minipage}[b]{0.3\textwidth}
        \centering
        \includegraphics[height=4cm]{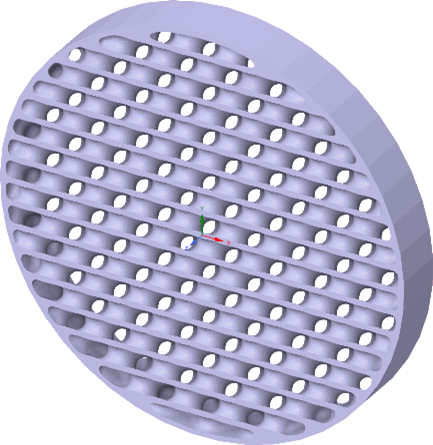}
    \end{minipage}
    \hfill
    \begin{minipage}[b]{0.3\textwidth}
        \centering
        \includegraphics[height=4cm]{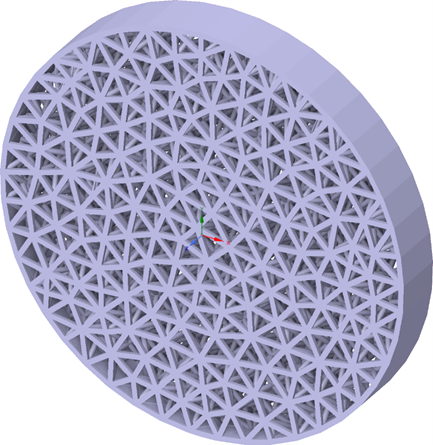}
    \end{minipage}

    \caption{The columns, arranged from left to right, present illustrations of the Gyroid and Schwarz D periodic minimal surface approximations, as well as a "Double Pyramid" lattice. The upper row depicts the fundamental structures, while the lower row showcases the simulated design specimens.}
    \label{fig:geometry_illustration}
\end{figure}

To mitigate direct wide-angular backscatter from curved structures, we generate a planar cap layer. The structures and specimens are depicted in Fig.~\ref{fig:geometry_illustration}. We utilise the following formulas to approximate the minimal surfaces, as referenced in Eq.~\ref{eq:minsurf1} for Gyroid minimal surfaces and Eq.~\ref{eq:minsurf2} for Schwarz D minimal surface:

\begin{equation}
    \sin(x)\cos(y) + \sin(y)\cos(z) + \sin(z)\cos(x) = 0\text{,}
    \label{eq:minsurf1}
\end{equation}   
\begin{equation}
    \cos(x)\cos(y)\cos(z) - \sin(x)\sin(y)\sin(z) = 0\text{.}
    \label{eq:minsurf2}
\end{equation}

The ability to model such structures using parametric, implicit surfaces offers significant advantages in terms of modelling efficiency during optimisation processes. However, in this study, we have opted to generate explicit geometry examples characterised by an equal volumetric density across all specimens and a minimal feature size or width of 0.5 mm to ensure manufacturability.
    
Although quasi-stochastic lattices present challenges when it comes to implicit modelling, they offer distinct benefits, including an increased surface area and the potential for inhomogeneous lattice densities with comparably high local variation gradients.

The triangularly meshed bodies can be directly exported as stereo-lithography (.stl) files and manufactured using additive manufacturing processes, such as laser powder bed fusion. These meshes can also be utilised directly for optical simulations, eliminating the need for meshing in a pre-processing step. Depending on the application, a NURBS-based model can also be generated from the meshed geometry; however, this typically requires more memory. Following hypothetical manufacturing, the specimens can be coated with an absorbing layer, such as black anodised aluminium or absorbing chemical vapour deposition (CVD) coatings, among others.
    
For this study, we employ a simplified surface model that empirically approximates the performance of black anodised aluminium, featuring a partially Lambertian and partially Gaussian angular scatter distribution. A radiometrically highly accurate surface model is not crucial for this investigation, as we maintain consistency in the surface model across all specimens and focus solely on comparing the effects of the varied geometries. However, for true design optimisation, it is advisable to calibrate the simulations using bidirectional reflectance distribution function (BRDF) measurement files.
    
It is important to note the memory requirements associated with such explicitly modelled structures. Our reference point is a maximum-to-minimum feature ratio of 80 (e.g., 40 mm diameter / 0.5 mm structure width). If we increase this ratio to 160 by halving the structure’s width while maintaining a constant volumetric density, the required memory increases by a factor of seven to eight. Conversely, if we reduce the structure’s width by a factor of five, the memory requirement for the Schwarz D model escalates by approximately a factor of 103, while for the "Double Pyramid" lattice, it increases by about a factor of 141. This can pose significant challenges regarding memory usage, potentially requiring tens of gigabytes to several terabytes, even for moderately complex systems. For larger systems, it is advisable to model the macroscopic structural light absorbers using implicit geometry, such as computing the intersection with the geometry for each simulation ray, rather than explicitly modelling the full structures.

\section*{Results}
\label{sec:results}
\begin{figure}[htbp]
    \centering

    \begin{minipage}[b]{0.3\textwidth}
        \centering
        \includegraphics[width=\linewidth]{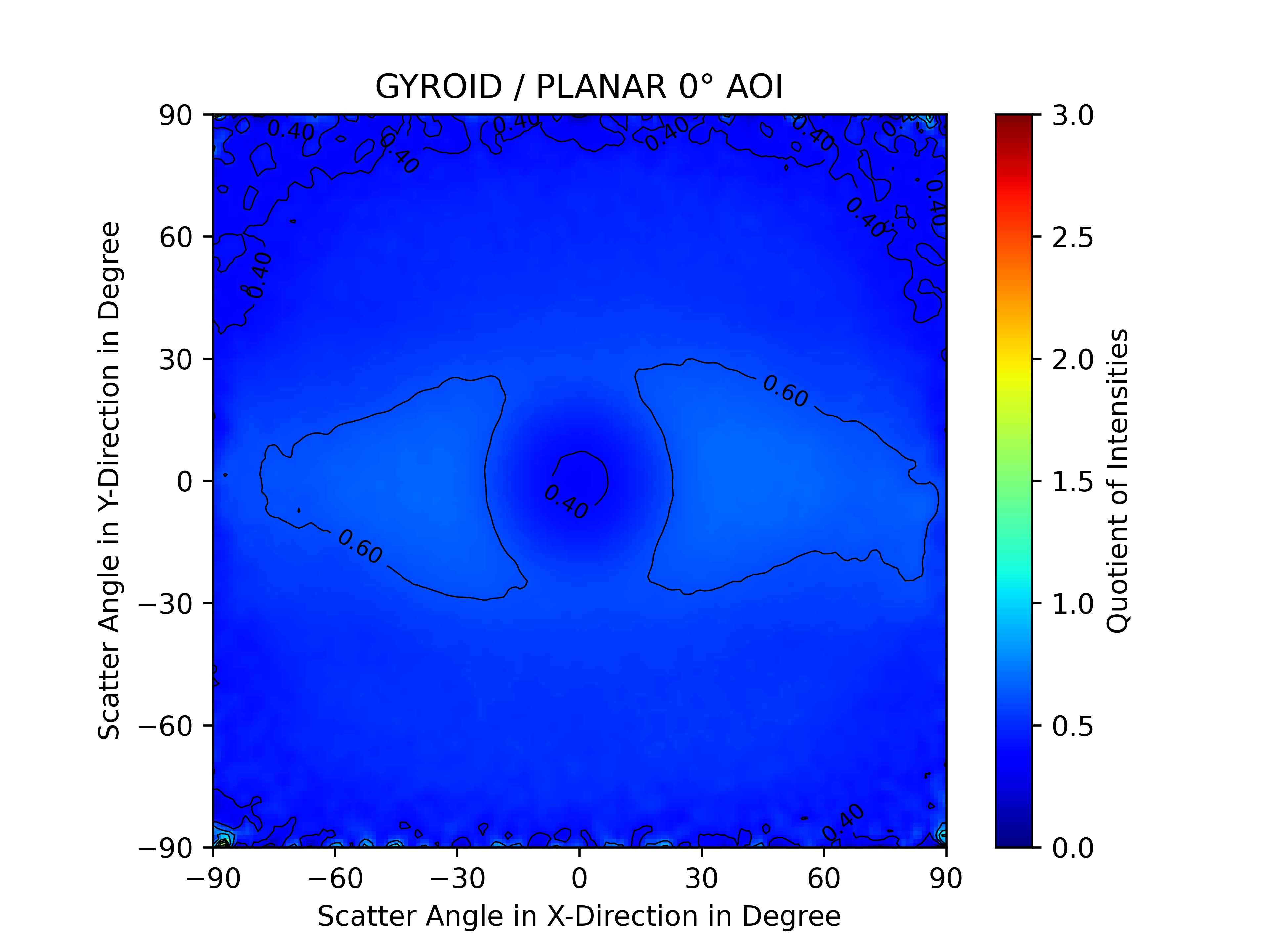}
    \end{minipage}
    \hfill
    \begin{minipage}[b]{0.3\textwidth}
        \centering
        \includegraphics[width=\linewidth]{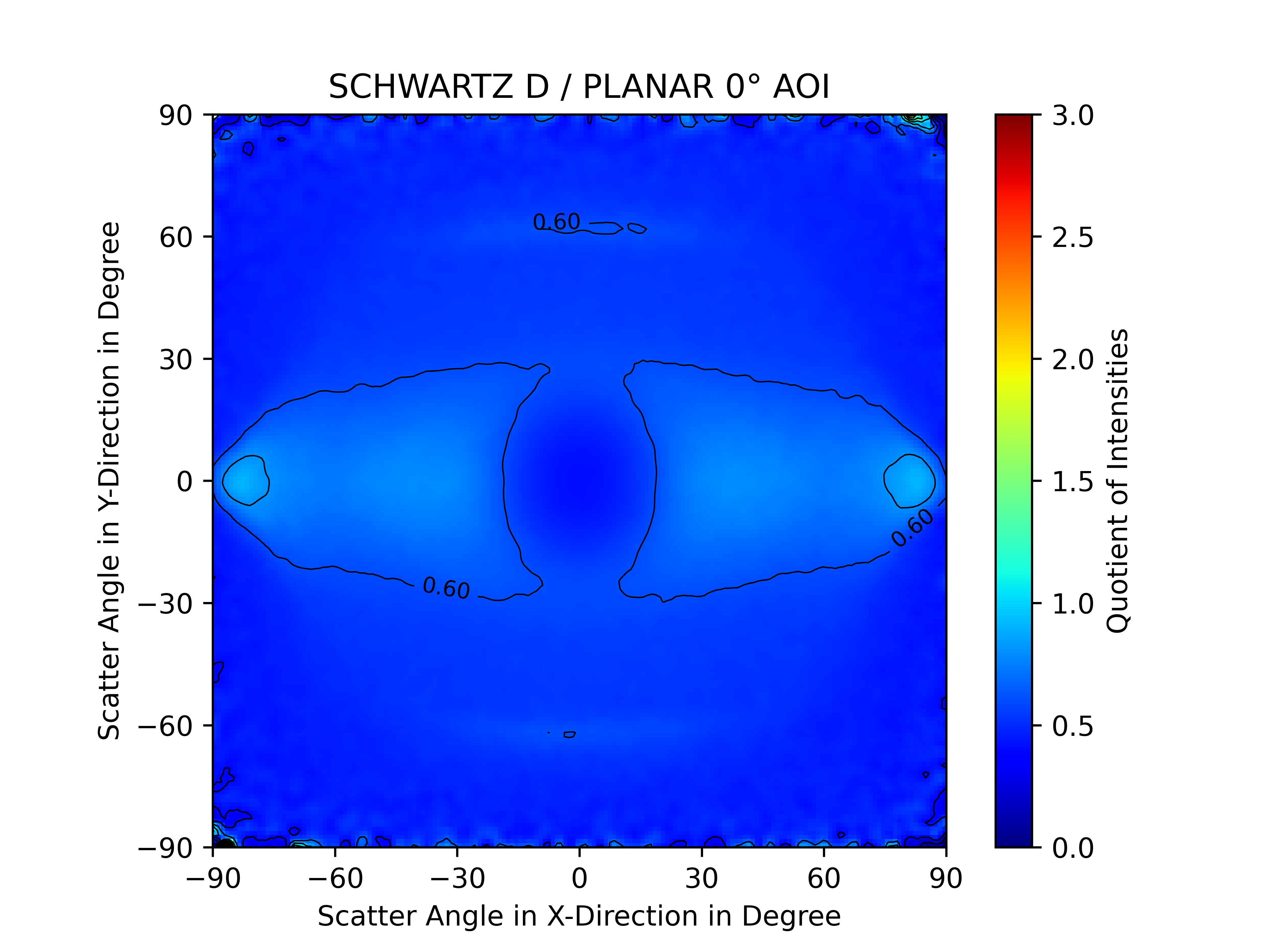}
    \end{minipage}
    \hfill
    \begin{minipage}[b]{0.3\textwidth}
        \centering
        \includegraphics[width=\linewidth]{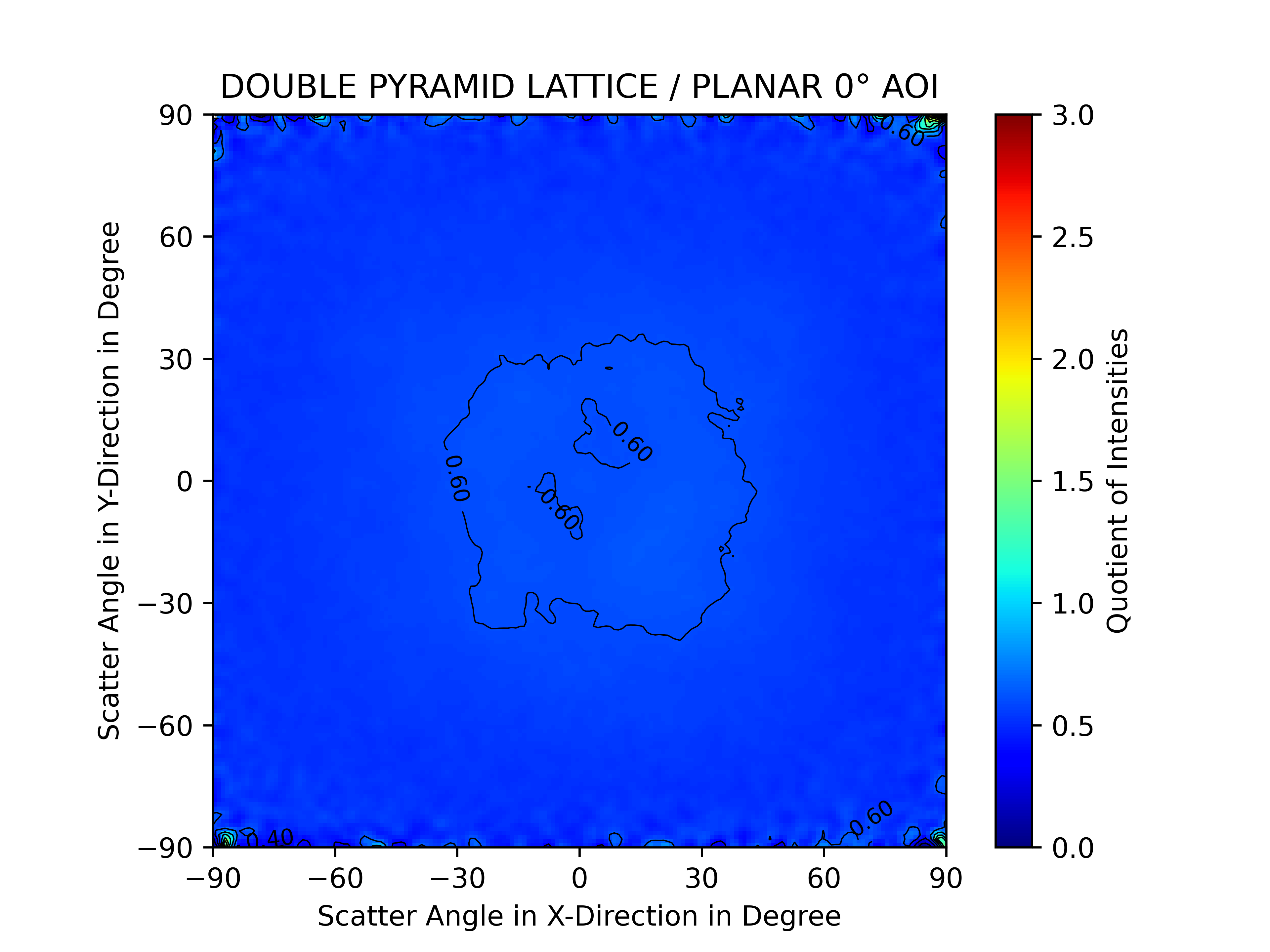}
    \end{minipage}

    \vspace{0.5cm} 

    \begin{minipage}[b]{0.3\textwidth}
        \centering
        \includegraphics[width=\linewidth]{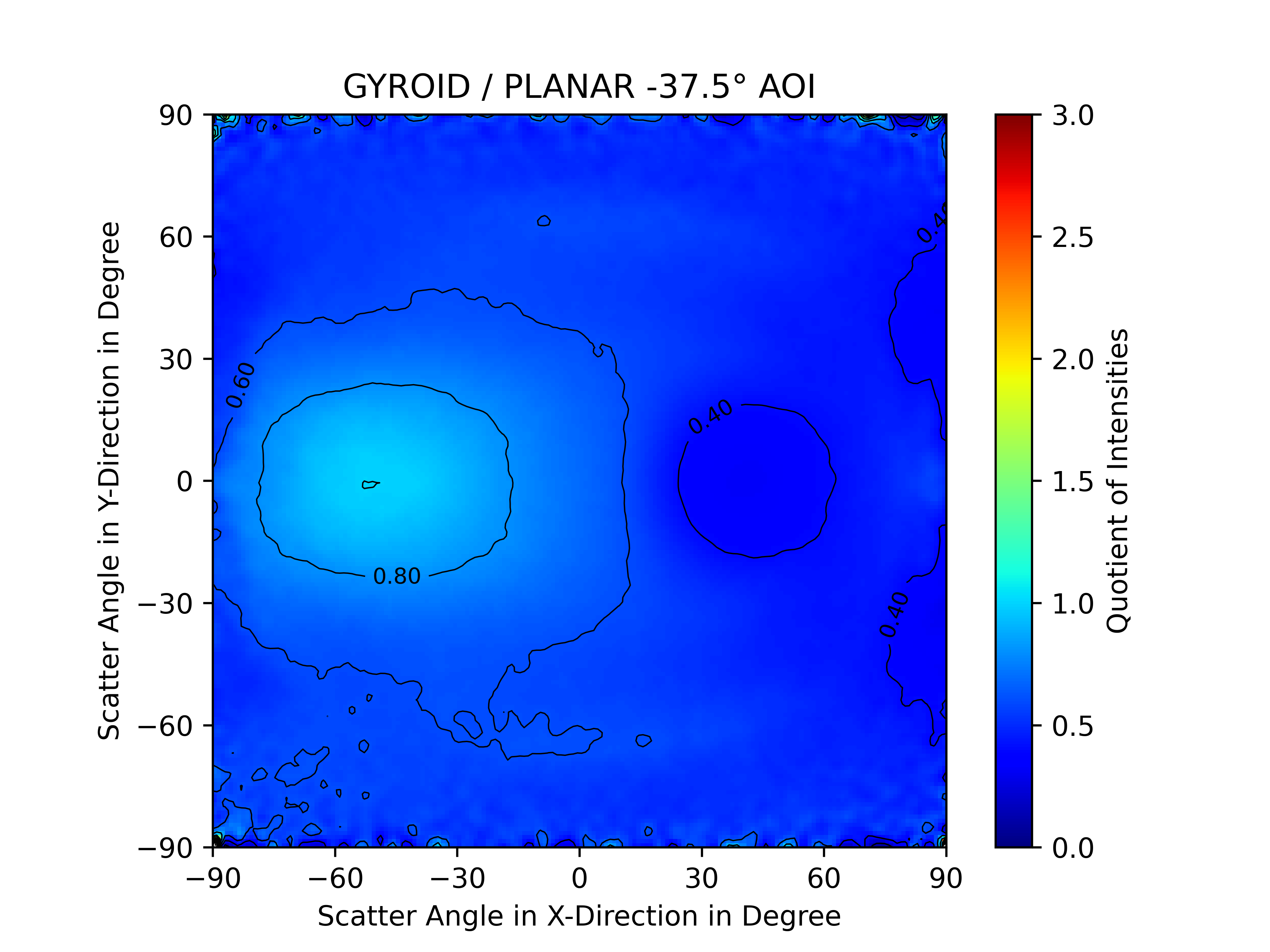}
    \end{minipage}
    \hfill
    \begin{minipage}[b]{0.3\textwidth}
        \centering
        \includegraphics[width=\linewidth]{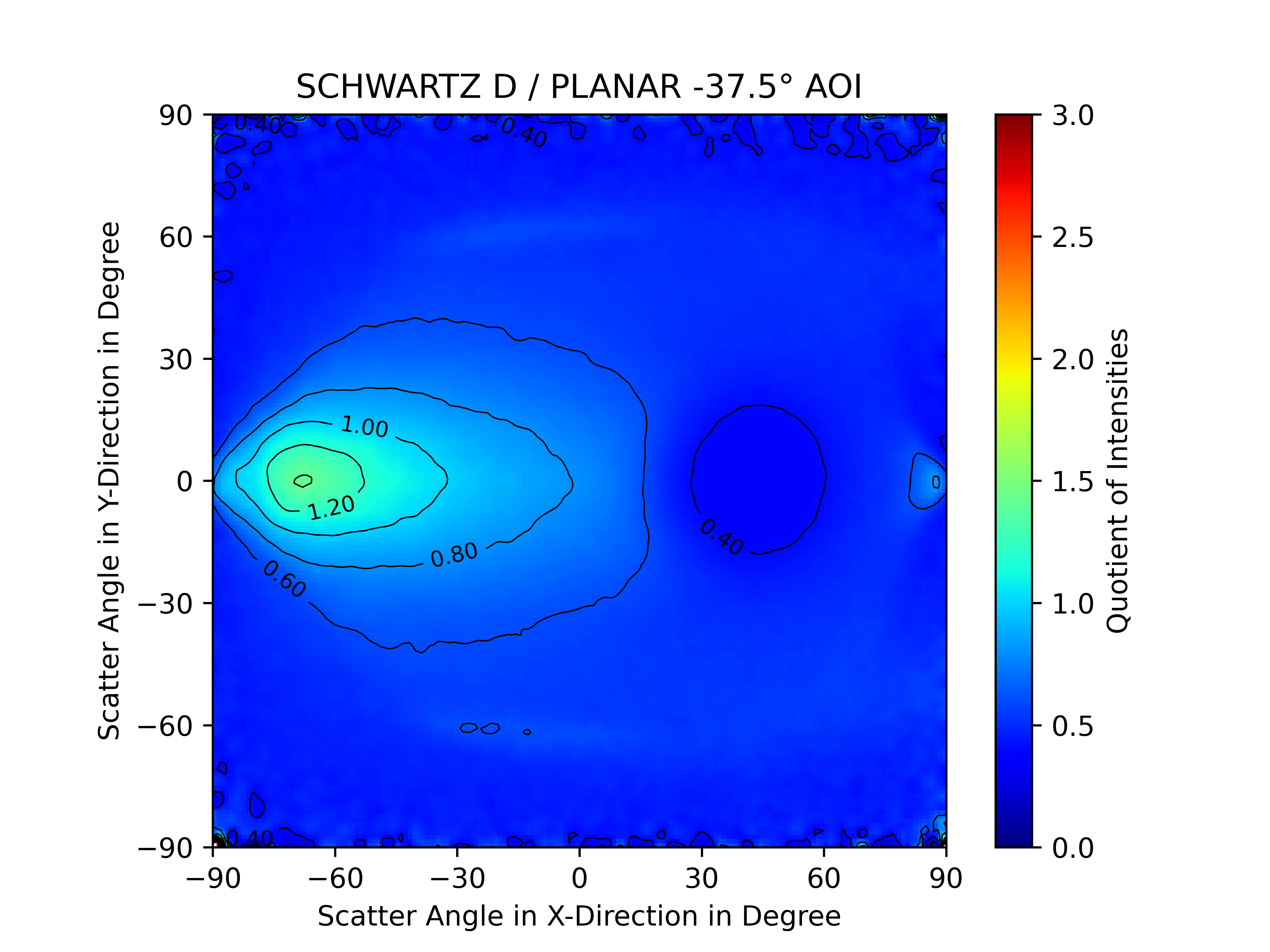}
    \end{minipage}
    \hfill
    \begin{minipage}[b]{0.3\textwidth}
        \centering
        \includegraphics[width=\linewidth]{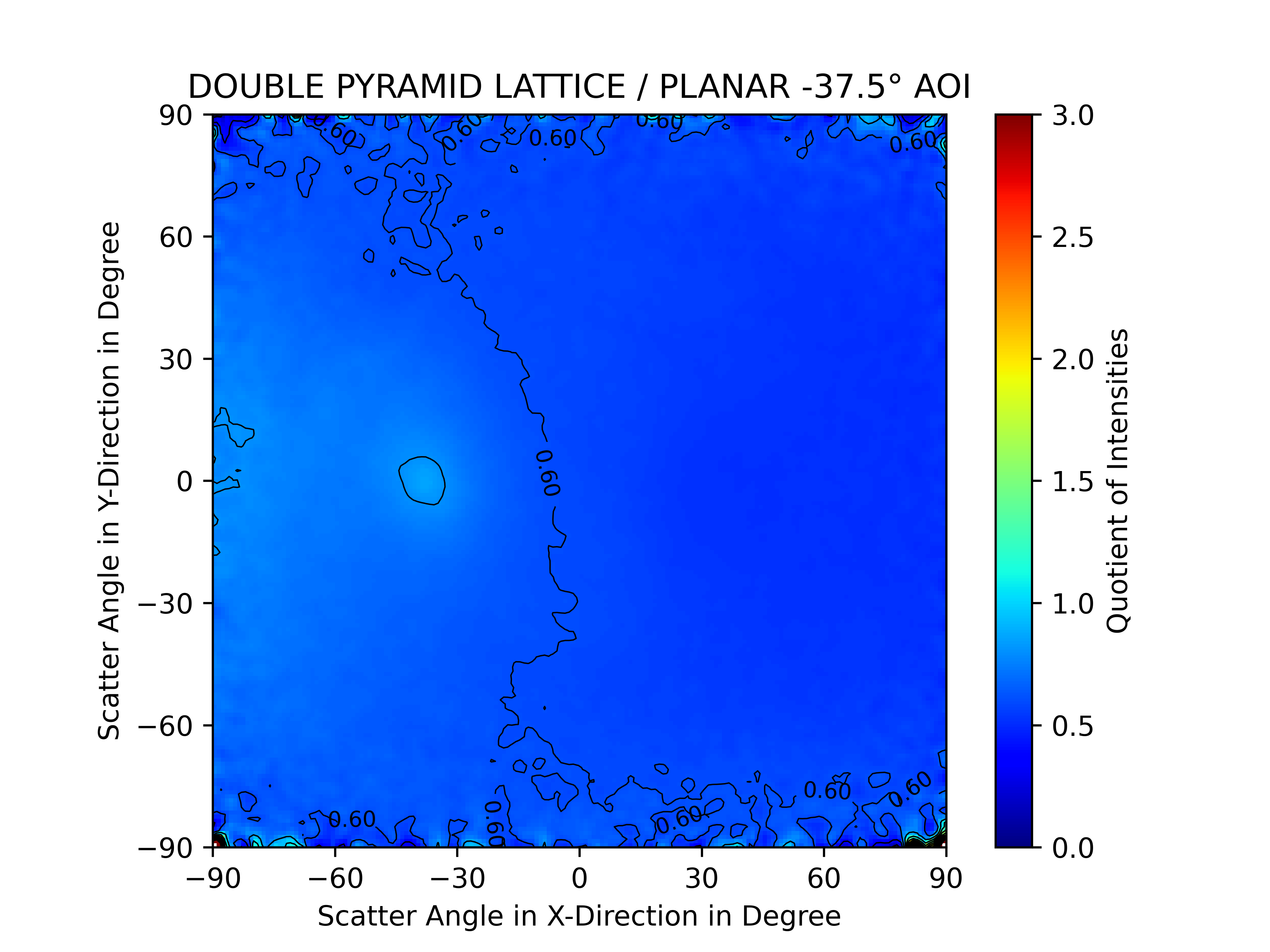}
    \end{minipage}

    \begin{minipage}[b]{0.3\textwidth}
        \centering
        \includegraphics[width=\linewidth]{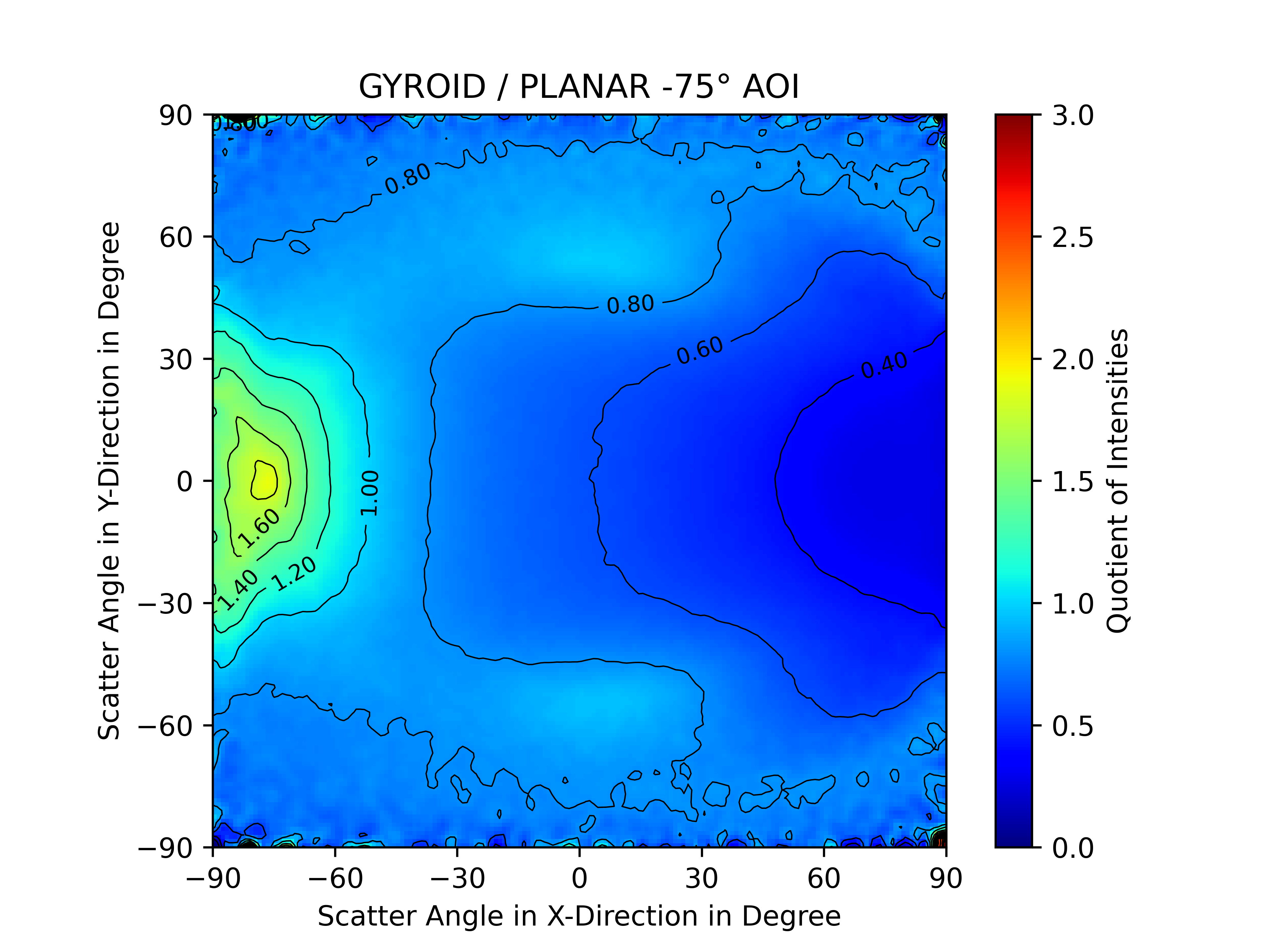}
    \end{minipage}
    \hfill
    \begin{minipage}[b]{0.3\textwidth}
        \centering
        \includegraphics[width=\linewidth]{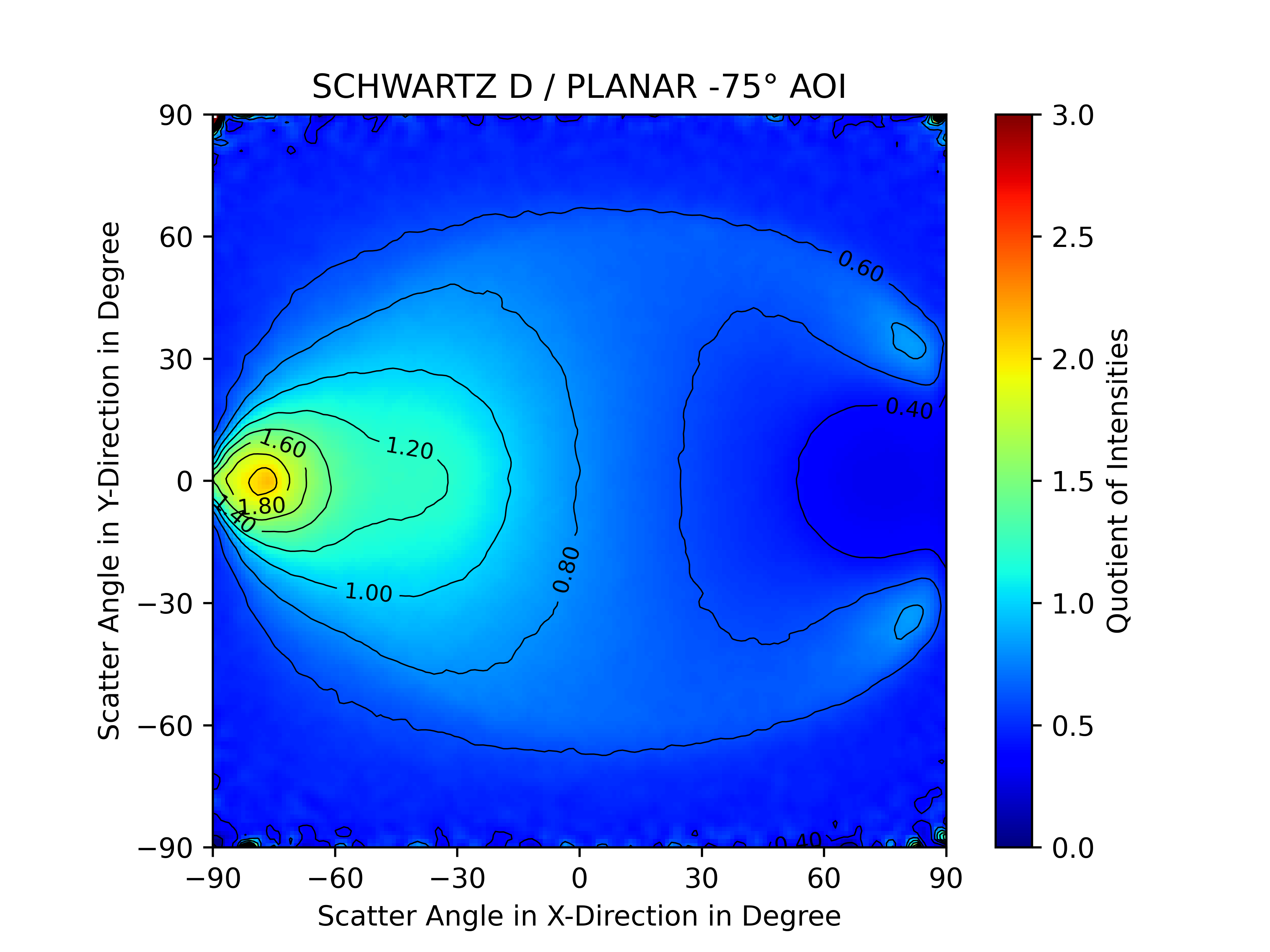}
    \end{minipage}
    \hfill
    \begin{minipage}[b]{0.3\textwidth}
        \centering
        \includegraphics[width=\linewidth]{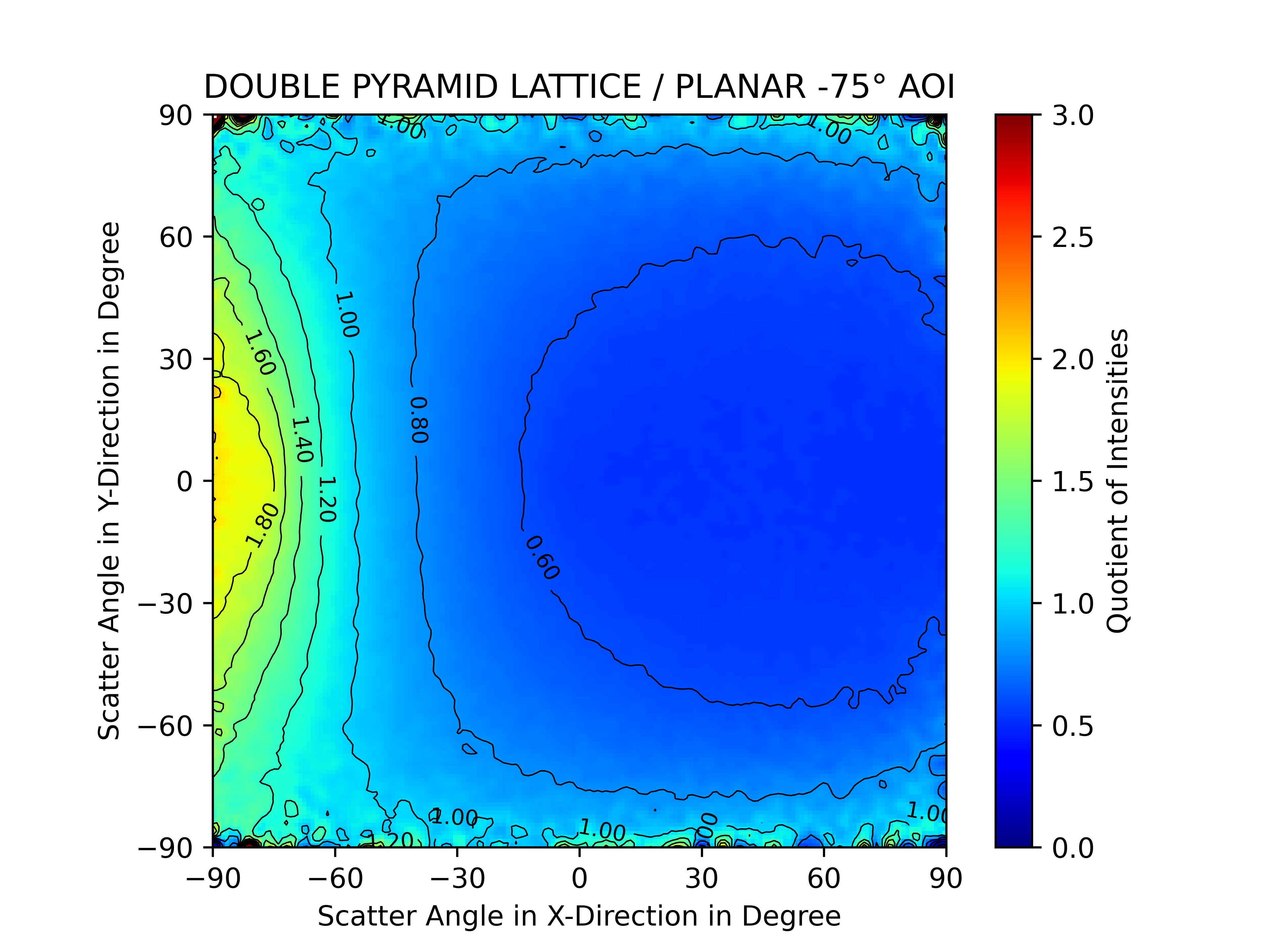}
    \end{minipage}

    \begin{minipage}[b]{0.3\textwidth}
        \centering
        \includegraphics[width=\linewidth]{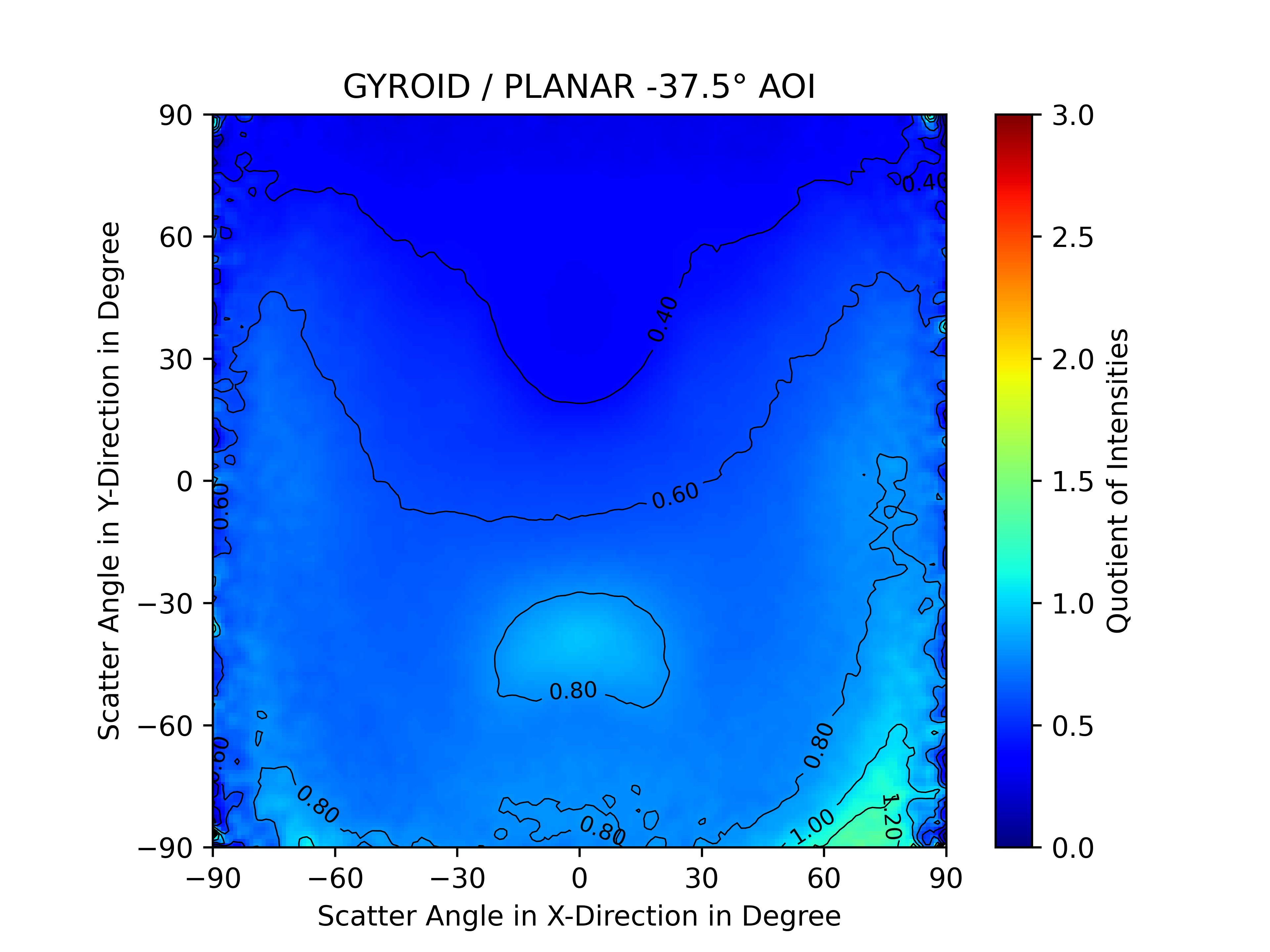}
    \end{minipage}
    \hfill
    \begin{minipage}[b]{0.3\textwidth}
        \centering
        \includegraphics[width=\linewidth]{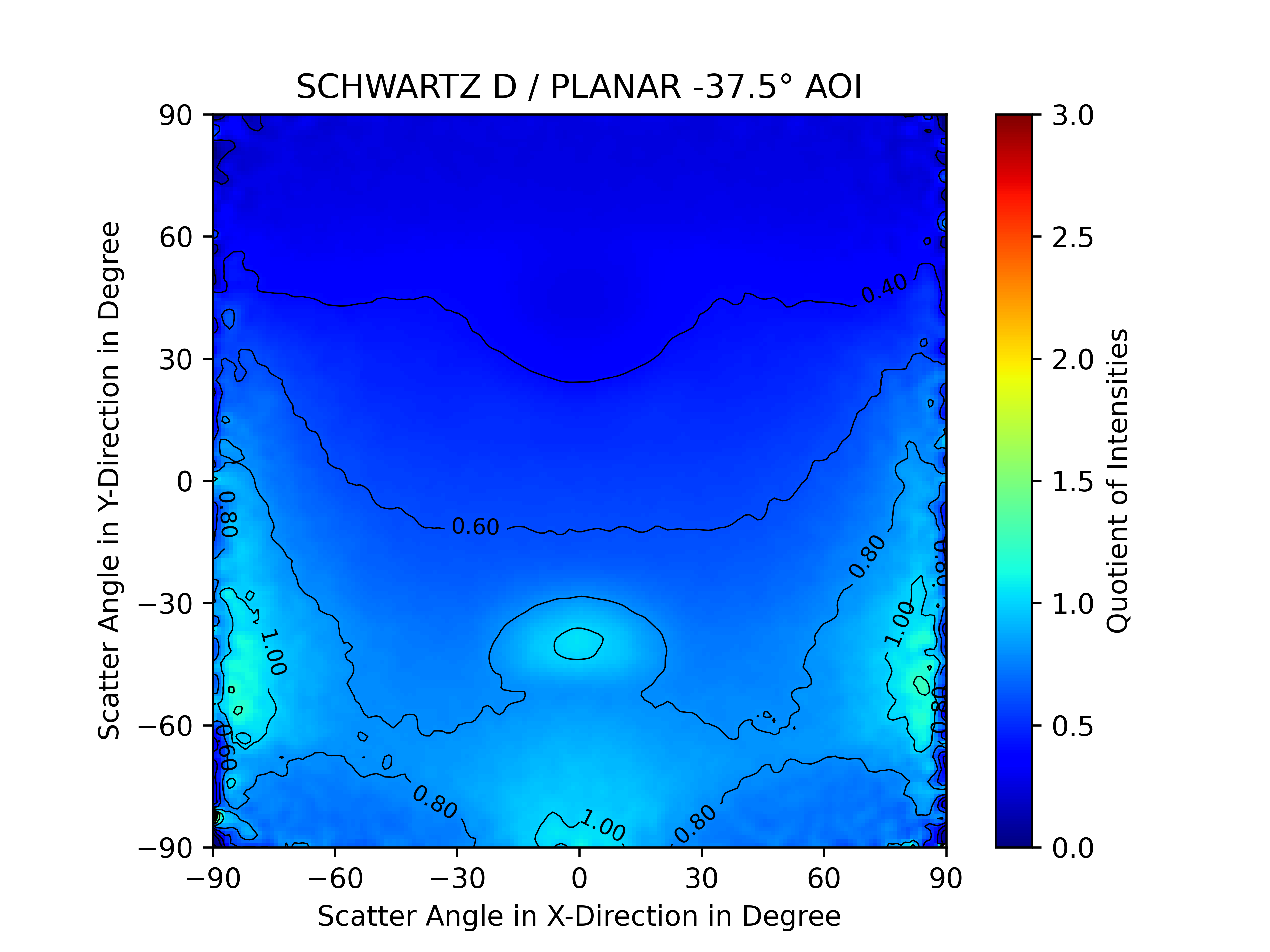}
    \end{minipage}
    \hfill
    \begin{minipage}[b]{0.3\textwidth}
        \centering
        \includegraphics[width=\linewidth]{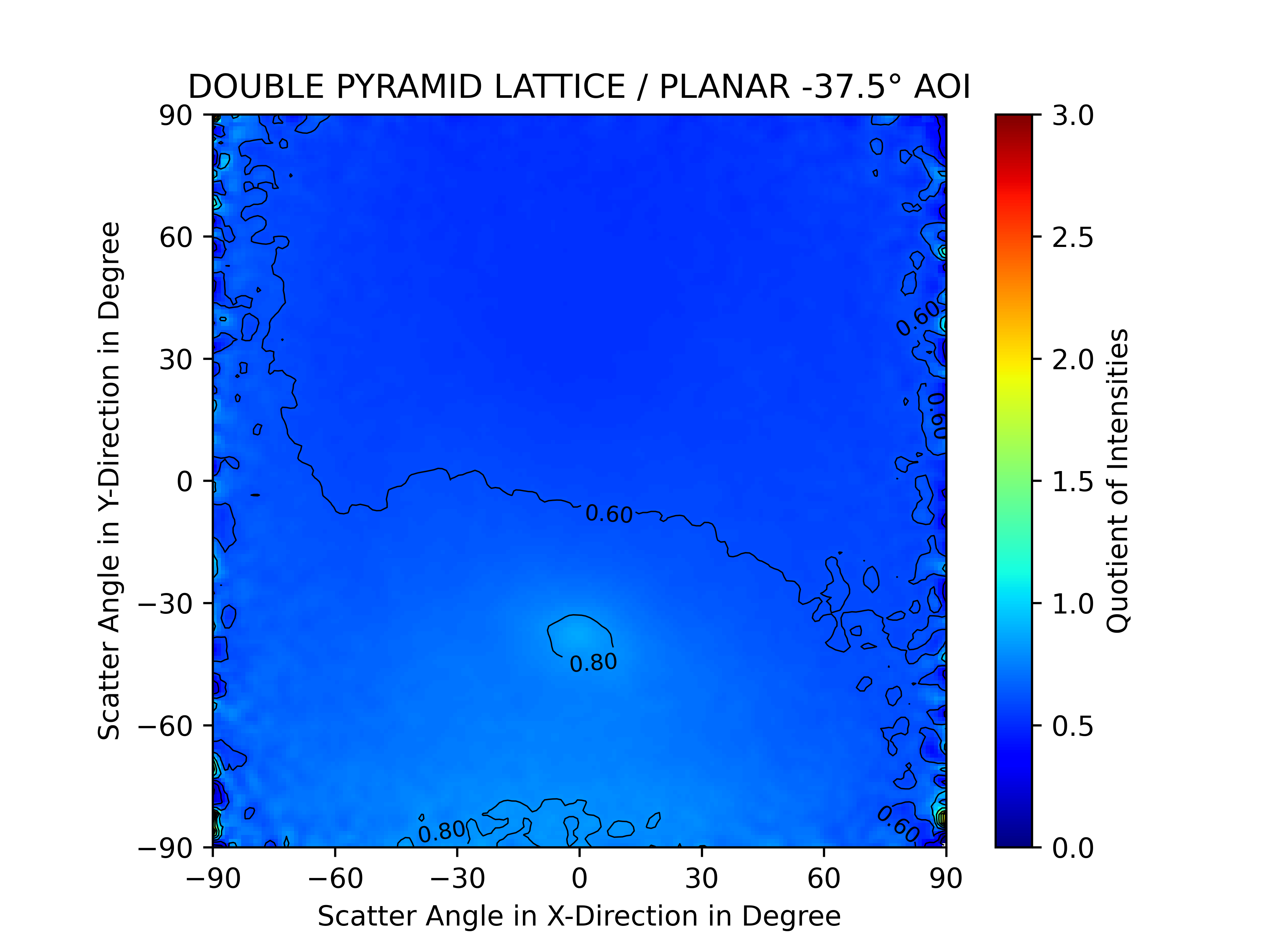}
    \end{minipage}

    \begin{minipage}[b]{0.3\textwidth}
        \centering
        \includegraphics[width=\linewidth]{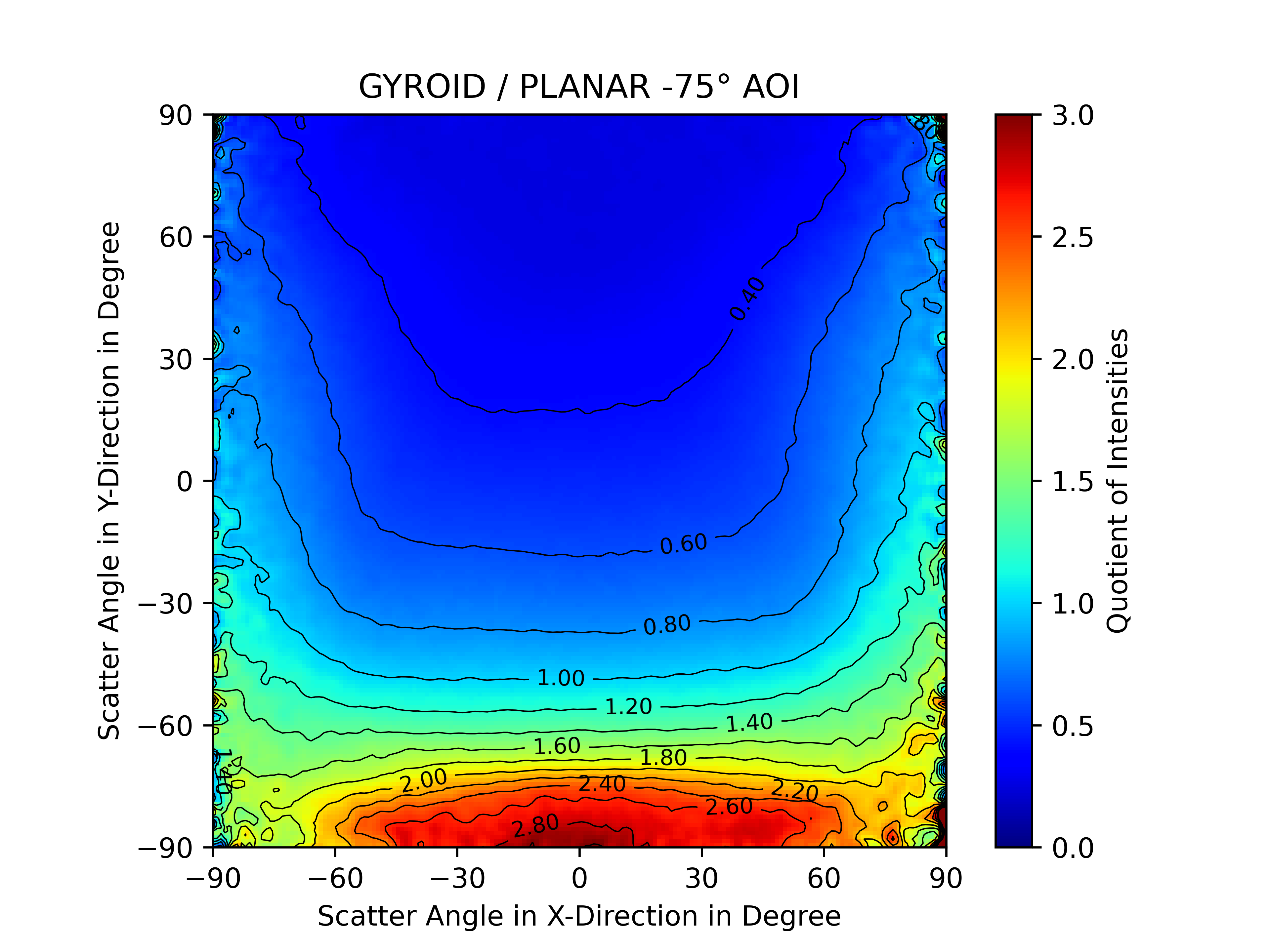}
    \end{minipage}
    \hfill
    \begin{minipage}[b]{0.3\textwidth}
        \centering
        \includegraphics[width=\linewidth]{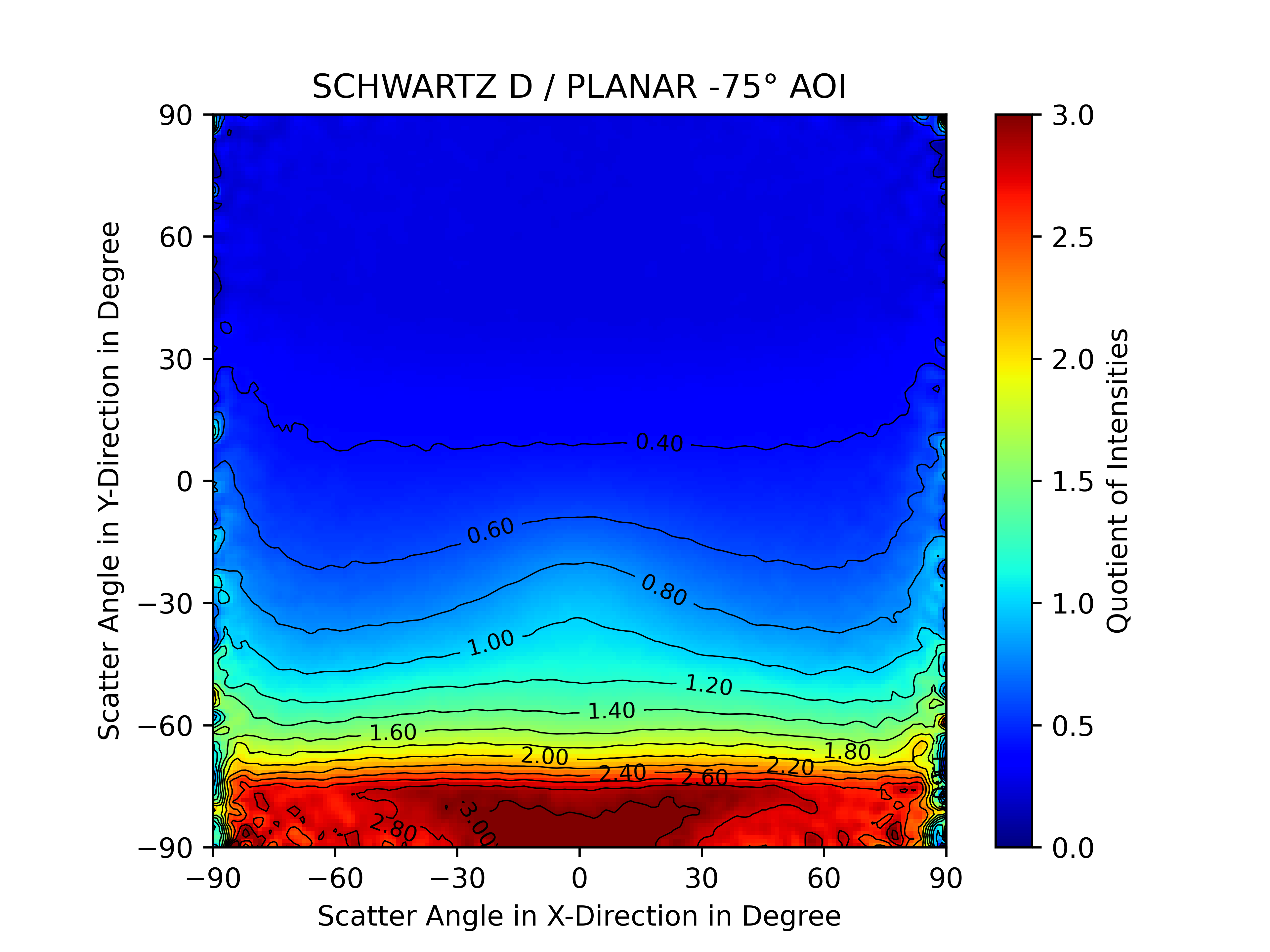}
    \end{minipage}
    \hfill
    \begin{minipage}[b]{0.3\textwidth}
        \centering
        \includegraphics[width=\linewidth]{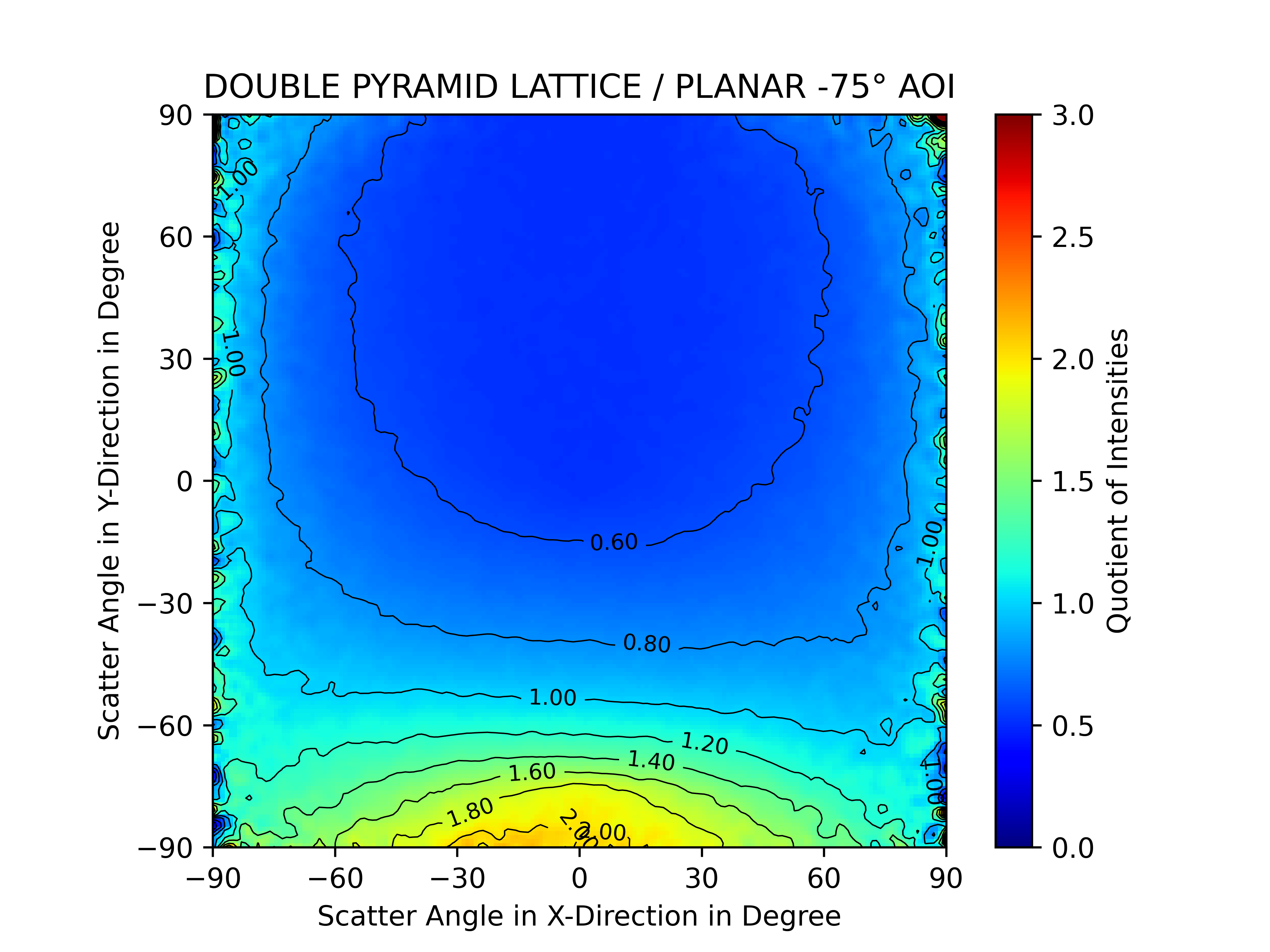}
    \end{minipage}

    \caption{From left to right, the columns represent the ratios of the reflected intensities for the Gyroid type specimen to the planar reference, the Schwarz D type specimen to the planar reference, and the "Double Pyramid" lattice type specimen to the planar reference. First row: Irradiance with normal incidence. Second (XZ-plane) and fourth (YZ-plane) rows: Irradiance at -37.5° to normal. Third (XZ-plane) and fifth (YZ-plane) rows: irradiance at -75 ° to normal. Z-axis is normal to the paper plane. The data from the hemispherical receiver is mapped onto a square in Cartesian coordinates for plotting.}
    \label{fig:results}
\end{figure}

Fig.~\ref{fig:results} (shown on page \pageref{fig:results}) presents the ratios of the reflected hemispherical intensity distributions for the investigated macroscopic structural light absorbers to the planar reference, respectively.

Table~\ref{tab:intensity_ratios} presents a results summary, for which we computed a ratio derived from the simulations of several specimens in comparison to a planar reference. 

\captionsetup[table]{skip=6pt}  
\begin{table}[htbp]
	\centering
	\begin{tabular}{|l|c|c|c|c|}
	\hline
	\textbf{Specimen / Reference} & \textbf{AOI} & \textbf{Max. Intensity Ratio} & \textbf{Avg. Intensity Ratio} \\
	 & & $Max(\mu_{\text{ratio}}) \pm 3\epsilon_{\text{ratio}}$ & $Avg(\mu_{\text{ratio}}) \pm 3\epsilon_{\text{ratio}}$ \\
	\hline
	Gyroid / Planar & 0° & 0.382 $\pm$ 4.24E-03 & 0.544 $\pm$ 4.94E-05 \\
	                & XZ 37.5° & 0.328 $\pm$ 3.96E-03 & 0.576 $\pm$ 4.89E-05 \\
	                & XZ 75.0° & 0.275 $\pm$ 3.76E-03 & 0.645 $\pm$ 4.73E-05 \\
	                & YZ 37.5° & 0.319 $\pm$ 3.90E-03 & 0.553 $\pm$ 4.93E-05 \\
	                & YZ 75.0° & 0.259 $\pm$ 3.63E-03 & 0.600 $\pm$ 4.84E-05 \\
	\hline
	Schwarz D / Planar & 0° & 0.417 $\pm$ 4.50E-03 & 0.578 $\pm$ 4.89E-05 \\
	                   & XZ 37.5° & 0.345 $\pm$ 4.09E-03 & 0.591 $\pm$ 4.87E-05 \\
	                   & XZ 75.0° & 0.337 $\pm$ 4.27E-03 & 0.701 $\pm$ 4.52E-05 \\
	                   & YZ 37.5° & 0.311 $\pm$ 3.88E-03 & 0.540 $\pm$ 4.94E-05 \\
	                   & YZ 75.0° & 0.273 $\pm$ 3.72E-03 & 0.604 $\pm$ 4.83E-05 \\
	\hline
	"Double Pyramid" Lattice / Planar & 0° & 0.613 $\pm$ 5.82E-03 & 0.581 $\pm$ 4.92E-05 \\
                                & XZ 37.5° & 0.518 $\pm$ 5.37E-03 & 0.589 $\pm$ 4.90E-05 \\
                                & XZ 75.0° & 0.525 $\pm$ 5.69E-03 & 0.665 $\pm$ 4.69E-05 \\
                                & YZ 37.5° & 0.518 $\pm$ 5.36E-03 & 0.596 $\pm$ 4.89E-05 \\
                                & YZ 75.0° & 0.512 $\pm$ 5.56E-03 & 0.661 $\pm$ 4.70E-05 \\
	\hline
	\end{tabular}
	\caption{Enhancement of the maximum and average relative intensity ratios of the reflected light for structured specimens, referenced against a planar specimen with similar coatings (where a smaller value is preferable).}
	\label{tab:intensity_ratios}
\end{table}

For each receiver cell (or "pixel"), we calculated the mean $\mu$ and standard deviation $\sigma$ of the incident rays’ energies, which allows for the identification of the maximum intensity and its uncertainty. Additionally, these statistics were computed across all cells to determine the average intensity. The standard deviation was subsequently divided by the square root of the number of incident rays (N) within the corresponding cells to derive a standard error $\epsilon$, cf. Eq.~\ref{eq:std_err}.

\begin{equation}
	\epsilon = \frac{\sigma}{\sqrt{N}}
	\label{eq:std_err}
\end{equation}

The mean cell values were then divided to calculate a ratio $\mu_{\text{ratio}}$, such as the maximum and average intensity ratio; cf.\ Eq.~\ref{eq:mean}.

\begin{equation}
	\mu_{\text{ratio}} = \frac{\mu_{\text{specimen}}}{\mu_{\text{planar}}}
	\label{eq:mean}
\end{equation}

To ascertain the effective error of the ratio $\epsilon_{\text{ratio}}$, we employed the root-sum-squared propagation of uncertainty for division operations, cf. Eq.~\ref{eq:std_err_ratio}. For this, we assumed randomness and the absence of correlation among the primary errors of subsequent simulations, and a normal distribution without skewness of the received ray energies.

\begin{equation}
	\epsilon_{\text{ratio}} = \mu_{\text{ratio}} \cdot \sqrt{
	\left( \frac{\epsilon_{\text{specimen}}}{\mu_{\text{specimen}}} \right)^2 +
	\left( \frac{\epsilon_{\text{planar}}}{\mu_{\text{planar}}} \right)^2}
	\label{eq:std_err_ratio}
\end{equation}

\section*{Discussion}
\label{sec:discussion}

\subsection*{Interpretation of Results}
\label{sec:interpretation_of_results}
The ratio results $\mu_{\text{ratio}}$ presented in Table~\ref{tab:intensity_ratios} exhibit deviations from the potential null hypotheses ($H0: \mu_{\text{ratio}} = 1$) that fall within the range of approximately 66 to 205 standard errors. Consequently, we refrain from further calculations of p-values and acknowledge the presence of a significant difference. Macroscopic structural light absorbers effectively diminish both the reflected peak and average intensities.

Furthermore, the results in Fig.~\ref{fig:results} provide insight into the directional distribution by illustrating the reflected intensity ratios of the specimens relative to the planar reference. In the forward scattering direction (for example, as light would be reflected from a planar mirror), the reflected intensities are diminished due to the geometries of the exemplary macroscopic structural light absorbers. Conversely, in the backward direction, the reflected intensities may be pronounced. This phenomenon can be either advantageous or disadvantageous, depending on the specific optical system that is to be optimised and the positioning of the macroscopic structural absorber within it. For instance, in an optical system characterised by unidirectional sequential light propagation, back-propagating stray light necessitates at least one additional reflection to propagate again in the critical direction. Notably, at the entrance of the optical system, stray light may be effectively rejected without incurring detrimental effects.

\subsection*{Comparison with Previous Studies}
\label{sec:comparison_with_previous_studies}
Using surface structuring is a common stray light improvement technique. For instance, by applying a saw-tooth or honeycomb pattern. However, the authors are not aware of prior art related to the specific optical absorbers presented. The concept outlined in \cite{rui2021lens} was the closest publication identified. In the present article, the underlying concept is based on the imitation of porous structures observed in foams and sponges, alongside the increasing availability of additive manufacturing techniques for producing components of nearly arbitrary complexity.

\subsection*{Limitations of this Study}
\label{sec:limitations_of_this_study}
In the present investigation, we limited the angular scan of the luminaire to normal incidence and two angles in orthogonal planes (XZ and YZ). To further develop this approach, it is advisable to conduct the luminaire scan in much smaller increments, such as $1^\circ$ or less, and to measure the hemispherical radiance rather than just the intensity. For instance, this could involve scanning a radiance detector with a small field of view over the hemisphere for each scan angle of the luminaire. It may also be necessary to analyse optical spectra instead of relying on single reference wavelengths. Additionally, depending on the radiometric surface properties, polarisation effects may not be negligible. A generalized approach would generate a substantial data set, which could be reduced by fitting parametric BRDF models to it; however, this may pose challenges for anisotropic BRDFs.

In the optical simulations, the surface topography was modelled as idealised isotropic scattering. While this approach is sufficient for the analysis at hand, anisotropic scattering is more commonly observed in abrasive machined surfaces or surfaces resulting from additive manufacturing (layer-wise build-up). Anisotropy can be reduced through processes such as orientation optimisation for additive manufacturing or post-processing techniques like glass bead blasting or wet chemical etching; however, it is rarely fully eliminated.

For stray light optimisation, it is crucial to ensure that there are no unintended geometric sources generating wide-angular scatter or tightly focused irradiances on critical locations such as optics or detector surfaces. To achieve this, it is important to characterise the radiometric properties of the surface through measurements (e.g., BRDF, reflectivity, absorbance, total integrated scatter, etc.) and to calibrate the simulations with this data. Such calibration is essential for ensuring sufficient predictive quality of the simulations and optimisation results.

The current investigation focuses solely on optical analysis; therefore, we modelled the bodies as surface shells to reduce memory requirements. In practical applications, structural mechanical requirements become relevant, especially if the stiffness of the structured part is critical for the optical system’s performance. In such cases, it is advisable to model with volumetric meshes that can be directly exported for structural mechanics analysis, such as finite element analysis, to assess stiffness, strength, elasticity, and other performance indicators.

This consideration also extends to the manufacturing of parts for physical measurements. While we prepared the samples to be directly manufacturable after adding a volumetric mesh, we did not produce samples and, consequently, did not measure their performance in physical experiments. Validation of the simulations through measurements and optimisation of the simulations via calibration for specific optical system designs has not yet been conducted.

Regarding the statistical analysis, the root-sum-squared method we applied to estimate error propagation is a simplified approximation that relies on numerous assumptions. A more general approach could be employed; however, for the purposes of our investigation and due to the low statistical dispersion determined so far, this seemed to be an unnecessary effort that would not significantly affect the interpretation of the results.

Given that we traced a finite number of rays to a pixelated receiver to compute the radiometric statistics, the noise present in the pixelated images can be characterised as shot noise. For instance, the computational noise is inversely related to the square root of the number of incident rays on each pixel. However, this should not be conflated with the statistical distribution of ray energy values within an individual cell. Unfortunately, with the commercial simulation software, we do not have access to evaluate the statistical distribution of ray energy values for individual cells, only to the mean and standard error values.

\subsection*{Future Directions}
\label{sec:future_directions}
In this proof-of-concept investigation, we demonstrated the principle by generating simple specimens from discs in Cartesian coordinates. This approach is suitable for applications in optical systems that lack symmetry, such as state-of-the-art lithography, laser, or AR/VR systems. However, for many classical optical systems, such as those used in microscopy or cinematography, cylindrical geometry is prevalent. Therefore, translating the minimal surface approximation into cylindrical coordinates would be a beneficial option.

Additional desirable design features would include local variations in the structure's geometry, encompassing the size of openings, local body widths, and the cross-sections and orientations of lattice struts. \cite{ibhadode2023topology} These variations should incorporate continuous transitions, designed through optimisation based on optical and structural simulations.

It is advantageous to model a macroscopic structural absorber as implicit geometry to facilitate rapid optimisation with reasonable memory requirements, in contrast to explicit geometry modelling. One approach to achieve this is by creating a user-defined feature for optical simulation software. Such user-defined features correlate input locations and angles of rays with output locations and angles of these rays, effectively serving as a parametric surrogate model of the macroscopic structural absorber. To generate this surrogate model efficiently, while also considering secondary objectives such as structural mechanical performance and weight reduction, we intend to employ probabilistic machine learning with adaptive design-of-experiments to generate training data. This will enable the creation of a regression model from which a macroscopic structural absorber can be retrieved and refined considering all relevant engineering physics effects following an initial hypervolume optimisation with the surrogate model.

In recent years, implicit geometry modelling has gained popularity in generative design and design-for-additive-manufacturing within mechanical design software. \cite{ntop2025}, \cite{altair2025inspire} These tools offer a range of engineering physics modelling options, including fluid dynamics, thermal and structural computations, as well as the optimisation of lattices and surfaces with this; however, to our knowledge, optics has not yet been integrated. From the author's perspective, incorporating optical ray tracing into such tools would be a valuable extension.

While periodic minimal surfaces can be modelled implicitly with relative efficiency, this is not necessarily the case for quasi-stochastic lattices that lack symmetry or periodicity. To further optimise macroscopic structural light absorbers using a generalised method, we anticipate exploring several approaches. For instance, to establish a first-order target for the reflected intensity, we can apply a light source mapped to a surface in the area of interest. Beginning with a Lambertian intensity, we will optimise the directed emittance to minimise detrimental stray light effects within the optical system. This intensity distribution will then serve as a target for generating a macroscopic structural absorber, potentially employing generative AI diffusion models or the like. \cite{maze2023diffusion} The desired topology will be created by eroding a volumetric section.

\subsection*{Conclusion}
\label{sec:conclusion}
In this investigation, we have demonstrated that the intensity of reflected stray light can be significantly reduced through the application of macroscopic structural light absorbers. By increasing the number of reflections before residual stray light propagates further into the optical system, we achieve reductions in peak stray light intensities by factors less than 0.39 and average stray light intensities by factors less than 0.65, without altering surface properties. Furthermore, the direction of scattering can be manipulated as a trade-off in optical performance, contingent upon the selection of structure and its specific implementation. For example, while forward scattering is diminished, this reduction may occur at the expense of increased backward scattering. This phenomenon is particularly relevant in optical systems characterised by unidirectional light transport, where the impact of backward scattering is less pronounced, as it necessitates an additional reflection to re-enter the critical propagation direction.
Macroscopic structural light absorbers may offer an intriguing trade-off between size and weight reduction, stiffness engineering, heat exchange and stray light performance, while also improving cost-effectiveness and the selection of durable coatings. This approach could be especially advantageous for aerospace applications.
Ultimately, we wish to emphasise that the macroscopic structures presented are not confined to stray light optimisation. They may also be relevant for thermal radiation absorbers or emitters. For instance, this can be achieved in conjunction with convective heat exchange through fluid or gas guiding channels incorporated within or through the structure. By applying highly reflective coatings instead of absorbing ones, these structures can also be employed for irradiance and intensity homogenisation, as required for integrating spheres and other non-imaging optical applications. 

\section*{Methods}
\label{sec:methods}

\subsection*{Specimens}
\label{sec:specimens}
A commercially available software tool was utilised to generate the geometries. \cite{ansys2025spaceclaim}

For all specimens, the volumetric densities of the resulting parts are maintained at a constant arbitrary value of 30.6\%. Additionally, we constrain the minimal feature width to 0.5 mm to ensure manufacturability using common additive manufacturing processes. This results in a periodic length of 5 mm for the Gyroid-type sample, 6.25 mm for the Schwarz D type, and a length setting of 2.85 mm for the "Double Pyramid" lattice.

For this investigation, we focus exclusively on opaque samples and optical simulations. Consequently, it is efficient to model the bodies using triangular surface shell meshes. The file sizes of the evaluated specimens are 138 MB for the Gyroid type, 79 MB for the Schwarz D type, and 14.5 MB for the "Double Pyramid" lattice type. These sizes correlate with the required number of triangles to mesh the shell surfaces (1 356 902 for the Gyroid type, 774 034 for the Schwarz D type, and 145 352 for the "Double Pyramid" lattice). While the total surface areas of the minimal surface type specimens are relatively similar (approximately 8 596 mm\textsuperscript{2} for the Gyroid type compared to 8 480 mm\textsuperscript{2} for the Schwarz D type), the surface area of the lattice specimen is approximately 12 569 mm\textsuperscript{2}, making it about 1.46 to 1.48 times larger.

\subsection*{Optical Simulations}
\label{sec:optical_simulations}
To evaluate the optical performance of the specimens, a commercially available software tool was used. \cite{ansys2025speos}
 
For the optical simulations, we selected a wavelength of 555 nm and did not account for polarization. We consider this approach acceptable, as we did not model properties that are sensitive to wavelength or polarization. We posit that the simulations remain valid for optical absorber coatings that sufficiently absorb across the relevant wavelength spectrum. Should there be significant variations in absorbance and reflectance across this wavelength spectrum, such factors can be modelled, although this may result in increased computation time for a similar simulation noise contrast.

The computations were performed using central processing units (CPUs), as the task is sufficiently straightforward to be completed within a few tens of seconds on a mobile workstation PC’s CPU. For each scan position of the luminaire, we traced 100 million rays, while neglecting the presence of an ambient material (e.g., air). Ray tracing was performed non-sequentially in a "direct" mode, also referred to as forward mode. This approach entails that a ray exits a source and propagates individually through the optical system until it either reaches a receiver or is terminated by other criteria, such as a maximum number of impacts. By setting the number of maximum surface interactions to one hundred, we ensured that no rays were halted by maximum number of impacts and that no ray propagation errors occurred in any of the simulations.
 
We activated a "weight" option to account for a ray's energy dissipation due to absorption during surface interactions. With this option, a ray's energy evolves through weighting as it propagates through the optical system until it reaches the receiver. Therefore, in the absence of ray tracing errors and rays lost due to non-interaction with a specimen along its path, all emitted rays contribute to the receiver's result, ensuring comparable statistics across different specimens. The minimum energy percentage parameter for the weighting option was maintained at a standard setting of 0.5\%.
The meshing of the shell surfaces was automatically generated by the design software upon the creation of structures based on minimal surface approximations and lattices. These meshes could have been refined; however, we opted to retain the initially generated meshes, as they are sufficiently accurate in comparison to NURBS representations of the geometries. Our intention was to compare the initial mesh sizes (number of triangles).

The artificial radiometric surface properties were consistent across all simulations and specimens. These properties were defined using a mixed Lambertian and Gaussian intensity scattering model, alongside the assumption of an opaque volume. The absorbance was established at 95\%, while the reflectance was set at 5\%. Of this reflectance, 85\% was scattered according to a Lambertian intensity distribution, and 15\% was scattered according to a Gaussian intensity distribution. The full width at half maximum (FWHM) angle of the reflected Gaussian intensity distribution for collimated input rays was set to 25 degrees. These values are estimated from previous BRDF measurements of black anodised aluminium for optical system applications, along with corresponding numerical fits, and are deemed sufficient for the analysis at hand. It was important for us to consider a surface that exhibits scattering characteristics including specular, near-specular (Gaussian), and wide-angle (Lambertian) scatter, which closely resembles a real reference surface.

\begin{figure}[htbp]
    \centering
    \begin{minipage}[c]{0.19\textwidth}
        \centering
        \includegraphics[height=3cm]{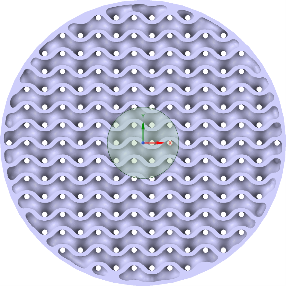}
    \end{minipage}
    \hfill
    \begin{minipage}[c]{0.19\textwidth}
        \centering
        \includegraphics[height=3cm]{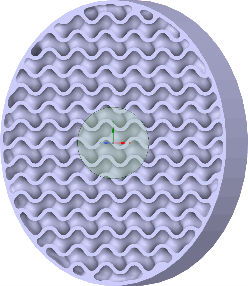}
    \end{minipage}
    \hfill
       \begin{minipage}[c]{0.19\textwidth}
        \centering
        \includegraphics[height=3cm]{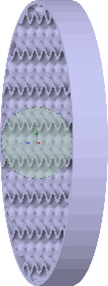}
    \end{minipage}
    \hfill
    \begin{minipage}[c]{0.19\textwidth}
        \centering
        \includegraphics[width=3cm]{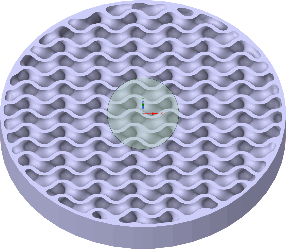}
    \end{minipage}
    \hfill 
    \begin{minipage}[c]{0.19\textwidth}
        \centering
        \includegraphics[width=3cm]{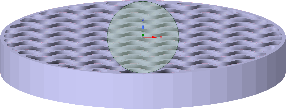}
    \end{minipage}
    \caption{Exemplary orthoscopic views for the Gyroid-type specimen are presented in directions normal to the five surface sources. The size and shape of each surface source are indicated by a green disc. From left to right: Normal incidence, 37.5\textdegree\ in XZ-plane, 75\textdegree\ in XZ-plane, 37.5\textdegree\ in YZ-plane, 75\textdegree\ in YZ-plane.}
    \label{fig:irradiance_illustration}
\end{figure}

The light sources were defined as light-emitting planar discs with a diameter of 10 mm. The radiant flux was set to 1 W, and the angular emission cone was defined as Lambertian, with a total angle of 0.01\textdegree\, effectively rendering the source a spatially uniform and collimated source. We did not alter the diameters or shapes of the sources with increasing angles of incidence to account for changes in the shapes and sizes of the irradiated spots; however, we ensured that a sufficient number of structures and structure periods were irradiated at any angle, and that no rays missed the specimen's surface for the largest angle of incidence. Angles of incidence were selected to be 0\textdegree\, 37.5\textdegree\, and 75\textdegree\ relative to the global z-axis, respectively, representing a minimal set to test the effects of normal, mid, and high angles of incidence. We tilted the source around the geometric centre of the specimen's cap layer, which was located at the origin of the global reference coordinate system, in two orthogonal planes (XZ and YZ). The specimens were aligned according to their symmetry (for those that exhibit symmetry). Fig.~\ref{fig:geometry_illustration} gives an impression of the symmetry and depicts the global origin as reference provides an impression of the symmetry and depicts the global origin as a reference. Fig.~\ref{fig:irradiance_illustration} illustrates the irradiated coverage of a specimen from the perspectives of the individual surface sources.

A radiometric hemispherical far-field intensity receiver is centred at the origin of the global coordinate system. The angular resolution is set to 1\textdegree\ in both the x- and y-directions. The mapping of the receiver cells is configured in spherical coordinates. This approach has the drawback that the cell sizes change significantly near the poles. A reduction in cell size correlates with increased statistical noise, as fewer rays impact a smaller cell. To compensate for this, we rotated the receiver in the azimuthal direction around the global z-axis, ensuring that the pole axis remains orthogonal to the plane of incidence for each simulation. An alternative option would have been to employ a conoscopic mapping. The data from the hemispherical receiver is mapped onto a square in Cartesian coordinates for further analysis and to illustrate the results.

To assess the radiometric accuracy of the results, we computed a relative standard error $\epsilon_{\text{relative}}$ for the maximum intensity for each simulation, respectively, and compared this value to a predetermined threshold, cf. Eq.~\ref{eq:rel_std_err} and Table~\ref{tab:rel_std_err}. In this proof-of-concept investigation, we accepted results as sufficiently radiometrically accurate for a comparative analysis if the peak error was less than five percent.

\begin{equation}
	\epsilon_{\text{relative}} = 100 \cdot \frac{\epsilon_{\text{specimen}}}{\mu_{\text{specimen}}}
	\label{eq:rel_std_err}
\end{equation}

\begin{table}[htbp]
	\centering
	\begin{tabular}{|l|c|c|}
	\hline
	\textbf{Specimen} & \textbf{AOI} & $\epsilon_{\text{relative}}$ in \% \\
	& & (for max. intensity ratios) \\
	\hline
	Gyroid & 0\textdegree & 0.942 \\
	       & XZ 37.5\textdegree & 1.046 \\
	       & XZ 75.0\textdegree & 1.211 \\
	       & YZ 37.5\textdegree & 1.062 \\
	       & YZ 75.0\textdegree & 1.250 \\
	\hline
	Schwarz D & 0\textdegree & 0.907 \\
	          & XZ 37.5\textdegree & 1.022 \\
	          & XZ 75.0\textdegree & 1.096 \\
	          & YZ 37.5\textdegree & 1.095 \\
	          & YZ 75.0\textdegree & 1.208 \\
	\hline
	"Double Pyramid" Lattice & 0\textdegree & 0.748 \\
	                         & XZ 37.5\textdegree & 0.843 \\
	                         & XZ 75.0\textdegree & 0.878 \\
	                         & YZ 37.5\textdegree & 0.841 \\
	                         & YZ 75.0\textdegree & 0.882 \\
	\hline
	Planar & 0\textdegree & 0.585 \\
	       & XZ 37.5\textdegree & 0.606 \\
	       & XZ 75.0\textdegree & 0.636 \\
	       & YZ 37.5\textdegree & 0.603 \\
	       & YZ 75.0\textdegree & 0.633 \\
	\hline
	\end{tabular}
	\caption{Relative standard errors for the simulated maximum intensities, from which the results are derived. The computed intensity distributions are considered sufficiently radiometrically accurate if the relative standard errors are less than five percent.}
	\label{tab:rel_std_err}
\end{table}

\subsection*{Data Analysis}
\label{sec:data_analysis}
The raw simulation data were analysed using features of commercial software and exported as text files, which were subsequently post-processed using custom Python programmes. To expedite the generation of the Python programmes, we employed Microsoft 365 Copilot as a coding assistant.

For the complete data range acquired by the hemispherical far-field intensity receiver, we calculated the maximum and average intensities, including the corresponding standard errors. We further processed this information by computing ratios to investigate the effect of macroscopic structural light absorbers. Additionally, we calculated ratios of the complete receiver datasets to elucidate the changes in the recorded intensity distributions caused by the specimen's structures in relation to a planar reference.

\subsection*{Manuscript}
\label{sec:manuscript}
To enhance the appeal of the originally drafted sections by integrating the stylistic nuances of British English, we employed a corporate-specific interface to GPT-4, hosted by Microsoft. For this, we used the prompt "Please revise the following text to conform to the standards of scientific British English: [Text]."

All responses generated by the large language model were meticulously reviewed for accuracy and integrity, and subsequently refined to align with the authors' intent.

\section*{Data Availability}
\label{sec:data_availability}
The data may be obtained from the corresponding author upon reasonable request.

\section*{Code Availability}
\label{sec:code_availability}
The python code used for the data analysis is available from the corresponding author upon reasonable request.

\section*{Acknowledgements}
\label{sec:acknowlegements}
This work was supported by research funding from the Design and Concepts Department of the Corporate Research and Technology Division at Carl Zeiss AG.

We thank Norbert Kerwien and Georgo Metalidis for their review and feedback prior to submitting the article.

\section*{Funding}
\label{sec:funding}
Software licences and computing hardware were provided by Carl Zeiss AG. The investigation and manuscript preparation were conducted during the author's winter holidays in 2024/2025.

\section*{Additional Information}
\label{sec:additional_information}
The manuscript underwent an internal review process within the organisation and received approval for publication to ensure that no classified information would be disclosed. The manuscript was finalised on 26 January 2025.

\section*{Ethics Declarations}
\label{sec:ethics_declarations}

\subsection*{Competing Interests}
\label{sec:competing_interests}
The authors do not declare any competing interests.

\subsection*{Use of Generative AI}
\label{sec:use_of_generative_ai}
Microsoft 365 Copilot and a corporate-specific interface to GPT-4, hosted by Microsoft, were utilised as coding assistants for data analysis and to refine the stylistic nuances of the manuscript to conform to British English standards.

All responses generated by the large language models were meticulously reviewed for accuracy and integrity, and subsequently refined to align with the authors' intent.

\bibliographystyle{IEEEtran}
\bibliography{references_v4}

\begin{thebibliography}{10}
\providecommand{\url}[1]{#1}
\csname url@samestyle\endcsname
\providecommand{\newblock}{\relax}
\providecommand{\bibinfo}[2]{#2}
\providecommand{\BIBentrySTDinterwordspacing}{\spaceskip=0pt\relax}
\providecommand{\BIBentryALTinterwordstretchfactor}{4}
\providecommand{\BIBentryALTinterwordspacing}{\spaceskip=\fontdimen2\font plus
\BIBentryALTinterwordstretchfactor\fontdimen3\font minus
  \fontdimen4\font\relax}
\providecommand{\BIBforeignlanguage}[2]{{%
\expandafter\ifx\csname l@#1\endcsname\relax
\typeout{** WARNING: IEEEtran.bst: No hyphenation pattern has been}%
\typeout{** loaded for the language `#1'. Using the pattern for}%
\typeout{** the default language instead.}%
\else
\language=\csname l@#1\endcsname
\fi
#2}}
\providecommand{\BIBdecl}{\relax}
\BIBdecl

\bibitem{harvey1977light}
J.~E. Harvey, ``Light-scattering characteristics of optical surfaces,'' in
  \emph{Proceedings of SPIE 0107, Stray Light Problems in Optical Systems},
  Reston, United States, 1977.

\bibitem{harvey2019understanding}
------, \emph{Understanding Surface Scatter: A Linear Systems
  Formulation}.\hskip 1em plus 0.5em minus 0.4em\relax Bellingham, United
  States: SPIE Press, 2019.

\bibitem{breault1977problems}
R.~P. Breault, ``Problems and techniques in stray radiation suppression,'' in
  \emph{Proceedings of SPIE 0107, Stray Light Problems in Optical Systems},
  Reston, United States, 1977.

\bibitem{fest2013straylight}
E.~Fest, \emph{Stray Light Analysis and Control}.\hskip 1em plus 0.5em minus
  0.4em\relax Bellingham, United States: SPIE Press, 2013.

\bibitem{clermont2024corot}
L.~Clermont, P.~Blain, W.~Khaddour, and W.~Uhring, ``Unlocking stray light
  mysteries in the {CoRot} baffle with the time-of-flight method,''
  \emph{Scientific Reports}, vol.~14, p. 6171, 2024.

\bibitem{clermont2024metop}
L.~Clermont, C.~Michel, Q.~Chouffart, and Y.~Zhao, ``Going beyond hardware
  limitations with advanced stray light calibration for the {Metop\textendash
  3MI} space instrument,'' \emph{Scientific Reports}, vol.~14, no.~1, p. 19490,
  2024.

\bibitem{chen2024deep}
M.~Chen, Y.~Zhao, W.~Yang, J.~Qian, S.~Li, Y.~Zheng, J.~Ma, S.~Wang, J.~Chen,
  and J.~Wei, ``A model for suppressing stray light in astronomical images
  based on deep learning,'' \emph{Scientific Reports}, vol.~14, no.~1, p.
  78472, 2024.

\bibitem{dobrowolski1995filters}
J.~A. Dobrowolski, L.~Li, and R.~A. Kemp, ``Metal/dielectric transmission
  interference filters with low reflectance. 1. design,'' \emph{Applied
  Optics}, vol.~34, no.~25, pp. 5673--5683, 1995.

\bibitem{cai2022absorber}
H.~Cai, S.~Shan, and X.~Wang, ``Optimal design of ultrabroadband
  omnidirectional planar structure absorber using anti-reflection coatings,''
  \emph{Journal of Optics}, vol.~51, pp. 154--160, 2022.

\bibitem{luhmann2020gold}
N.~Luhmann, D.~H{\o}j, M.~Piller, H.~K{\"a}hler, M.-H. Chien, R.~G. West, U.~L.
  Andersen, and S.~Schmid, ``Ultrathin 2\,nm gold as impedance-matched absorber
  for infrared light,'' \emph{Nature Communications}, vol.~11, p. 2161, 2020.

\bibitem{mizuno2009carbon}
K.~Mizuno, J.~Ishii, H.~Kishida, and K.~Hata, ``A black body absorber from
  vertically aligned single-walled carbon nanotubes,'' \emph{Proceedings of the
  National Academy of Sciences}, vol. 106, no.~15, pp. 6044--6047, 2009.

\bibitem{aydin2011plasmonic}
K.~Aydin, V.~E. Ferry, R.~M. Briggs, and H.~A. Atwater, ``Broadband
  polarization-independent resonant light absorption using ultrathin plasmonic
  super absorbers,'' \emph{Nature Communications}, vol.~2, p. 517, 2011.

\bibitem{he2024baffle}
W.~He, S.~Lin, Y.~Lai, D.~Ma, J.~Wang, Z.~Wang, X.~Zhang, and Y.~Jin, ``Review
  of the baffle technology application,'' in \emph{Proceedings of SPIE 13497,
  AOPC 2024: Optical Design and Manufacturing}, Beijing, China, 2024.

\bibitem{stavroudis1994baffles}
O.~N. Stavroudis and L.~D. Foo, ``System of reflective telescope baffles,''
  \emph{Optical Engineering}, vol.~33, no.~3, 1994.

\bibitem{mccoy2018feathers}
D.~E. McCoy, T.~Feo, T.~A. Harvey, and R.~O. Prum, ``Structural absorption by
  barbule microstructures of super black bird of paradise feathers,''
  \emph{Nature Communications}, vol.~9, p.~1, 2018.

\bibitem{vukusic2004butterfly}
P.~Vukusic, J.~R. Sambles, and C.~R. Lawrence, ``Structurally assisted
  blackness in butterfly scales,'' \emph{Proceedings of the Royal Society B:
  Biological Sciences}, vol. 271, no. Suppl 4, pp. S237--S239, 2004.

\bibitem{davis2020ultrablack}
A.~L. Davis, H.~F. Nijhout, and S.~Johnsen, ``Diverse nanostructures underlie
  thin ultra-black scales in butterflies,'' \emph{Nature Communications},
  vol.~11, p. 2704, 2020.

\bibitem{zhao2011carbon}
Q.~Zhao, T.~Fan, J.~Ding, D.~Zhang, Q.~Guo, and M.~Kamada, ``Super black and
  ultrathin amorphous carbon film inspired by anti-reflection architecture in
  butterfly wing,'' \emph{Carbon}, vol.~49, no.~3, pp. 877--883, 2011.

\bibitem{mackay1985minimal}
A.~L. Mackay, ``Periodic minimal surfaces,'' \emph{Nature}, vol. 314, pp.
  604--606, 1985.

\bibitem{fernandes2021lattices}
M.~C. Fernandes, J.~Aizenberg, J.~C. Weaver, and K.~Bertoldi, ``Mechanically
  robust lattices inspired by deep-sea glass sponges,'' \emph{Nature
  Materials}, vol.~20, no.~2, pp. 237--241, 2021.

\bibitem{prathyusha2022review}
A.~L.~R. Prathyusha and G.~R. Babu, ``A review on additive manufacturing and
  topology optimization process for weight reduction studies in various
  industrial applications,'' \emph{Materials Today: Proceedings}, vol.~62,
  no.~1, pp. 109--117, 2022.

\bibitem{daynes2025graded}
S.~Daynes and S.~Feih, ``Functionally graded lattice structures with tailored
  stiffness and energy absorption,'' \emph{International Journal of Mechanical
  Sciences}, vol. 285, 2025.

\bibitem{chen2018metamaterials}
D.~Chen and X.~Zheng, ``Multi-material additive manufacturing of metamaterials
  with giant, tailorable negative poisson's ratios,'' \emph{Scientific
  Reports}, vol.~8, p. 10880, 2018.

\bibitem{careri2023heat}
F.~Careri, R.~H.~U. Khan, C.~Todd, and M.~M. Attallah, ``Additive manufacturing
  of heat exchangers in aerospace applications: a review,'' \emph{Applied
  Thermal Engineering}, vol. 235, p. 121387, 2023.

\bibitem{wang2024tpms}
J.~Wang, C.~Qian, X.~Qiu, B.~Yu, L.~Yan, J.~Shi, and J.~Chen, ``Numerical and
  experimental investigation of additive manufactured heat exchanger using
  triply periodic minimal surfaces ({TPMS}),'' \emph{Thermal Science and
  Engineering Progress}, vol.~55, p. 103007, 2024.

\bibitem{an2022insulation}
L.~An, Z.~Guo, Z.~Li, Y.~Fu, Y.~Hu, Y.~Huang, F.~Yao, C.~Zhou, and S.~Ren,
  ``Tailoring thermal insulation architectures from additive manufacturing,''
  \emph{Nature Communications}, vol.~13, p. 4294, 2022.

\bibitem{ansys2025spaceclaim}
\BIBentryALTinterwordspacing
``{ANSYS} spaceclaim,'' accessed July 2025. [Online]. Available:
  \url{https://www.ansys.com/products/3d-design/ansys-spaceclaim}
\BIBentrySTDinterwordspacing

\bibitem{ansys2025speos}
\BIBentryALTinterwordspacing
``{ANSYS} {SPEOS},'' accessed July 2025. [Online]. Available:
  \url{https://www.ansys.com/products/optics/ansys-speos}
\BIBentrySTDinterwordspacing

\bibitem{schoen1970minimal}
\BIBentryALTinterwordspacing
A.~H. Schoen, ``Infinite periodic minimal surfaces without
  self-intersections,'' {NASA} Electronics Research Center, Technical Note
  {NASA{\textendash}TN{\textendash}D{\textendash}5541}, 1970. [Online].
  Available: \url{https://ntrs.nasa.gov/citations/19700020472}
\BIBentrySTDinterwordspacing

\bibitem{hoffman2001minimal}
\BIBentryALTinterwordspacing
D.~Hoffman, ``Computing minimal surfaces,'' in \emph{Global Theory of Minimal
  Surfaces. Proceedings of the Clay Mathematics Institute 2001 Summer
  School}.\hskip 1em plus 0.5em minus 0.4em\relax Berkeley, California: Clay
  Mathematics Institute, 2001. [Online]. Available:
  \url{https://www.claymath.org/resource/global-theory-of-minimal-surfaces/}
\BIBentrySTDinterwordspacing

\bibitem{rui2021lens}
\BIBentryALTinterwordspacing
S.~Rui, Z.~Zhuo, D.~Shuwu, W.~Ying, H.~Changning, L.~Chaolong, Z.~Xiaoming,
  D.~Lianqing, Y.~Lixin, and C.~Ying, ``Lens barrel structure suitable for
  bright background imaging condition,'' 2021. [Online]. Available:
  \url{https://patents.google.com/patent/CN114415312A/en}
\BIBentrySTDinterwordspacing

\bibitem{ibhadode2023topology}
O.~Ibhadode, Z.~Zhang, J.~Sixt, K.~M. Nsiempba, J.~Orakwe,
  A.~Martinez-Marchese, O.~Ero, S.~I. Shahabad, A.~Bonakdar, and E.~Toyserkani,
  ``Topology optimization for metal additive manufacturing: current trends,
  challenges, and future outlook,'' \emph{Virtual and Physical Prototyping},
  vol.~18, no.~1, p. e2181192, 2023.

\bibitem{ntop2025}
\BIBentryALTinterwordspacing
``{nTop},'' accessed July 2025. [Online]. Available:
  \url{https://www.ntop.com/}
\BIBentrySTDinterwordspacing

\bibitem{altair2025inspire}
\BIBentryALTinterwordspacing
``Altair inspire,'' accessed July 2025. [Online]. Available:
  \url{https://altair.com/inspire}
\BIBentrySTDinterwordspacing

\bibitem{maze2023diffusion}
F.~Maz{\'e} and F.~Ahmed, ``Diffusion models beat {GANs} on topology
  optimization,'' in \emph{Proceedings of the {AAAI} Conference on Artificial
  Intelligence}, vol.~37, no.~8, 2023, pp. 9108--9116.

\end{thebibliography}

%
%
%

\newpage
\section*{Supplementary Figures}
\label{sec:supplementary_figures}

\begin{figure}[htbp]
    \centering

    \begin{minipage}[b]{0.49\textwidth}
        \centering
        \includegraphics[width=\linewidth]{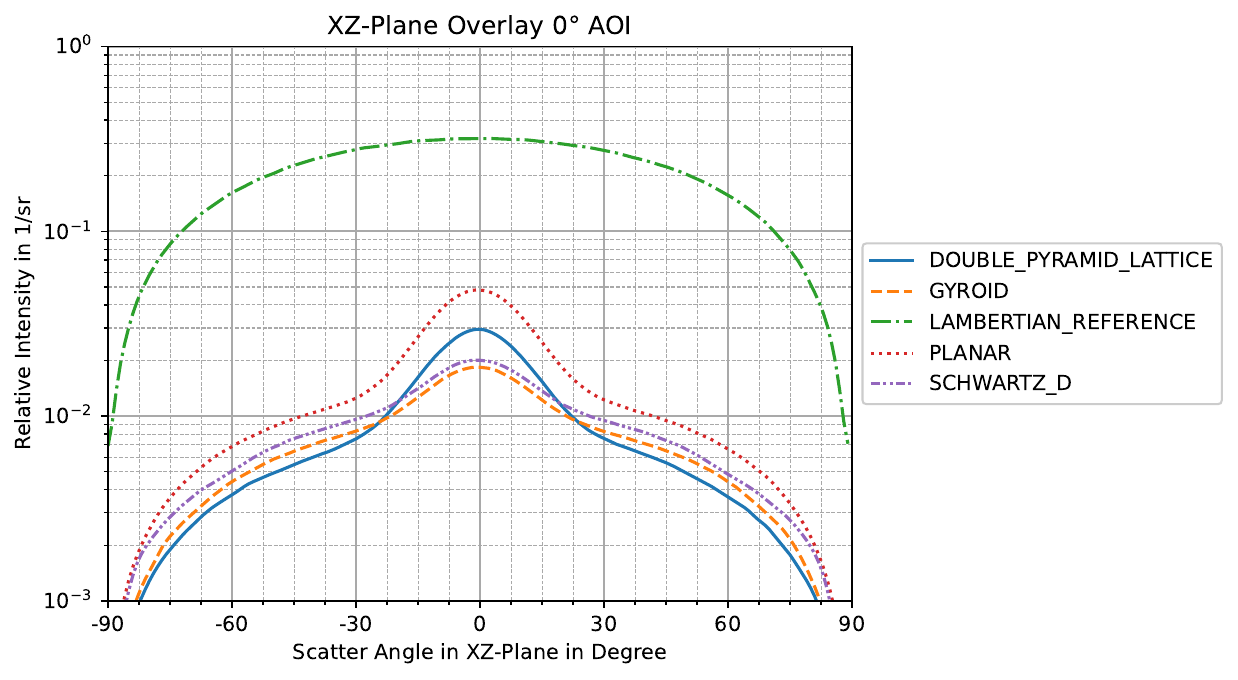}
    \end{minipage}
    \hfill
    \begin{minipage}[b]{0.49\textwidth}
        \centering
        \includegraphics[width=\linewidth]{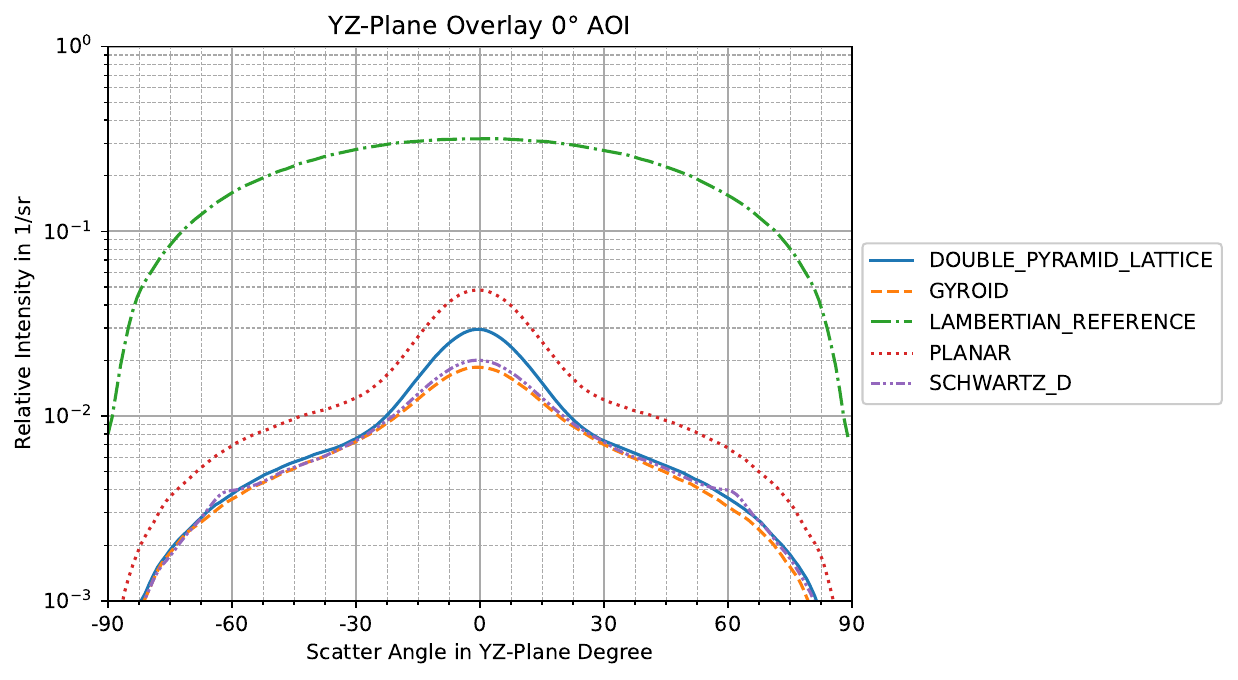}
    \end{minipage}

    \begin{minipage}[b]{0.49\textwidth}
        \centering
        \includegraphics[width=\linewidth]{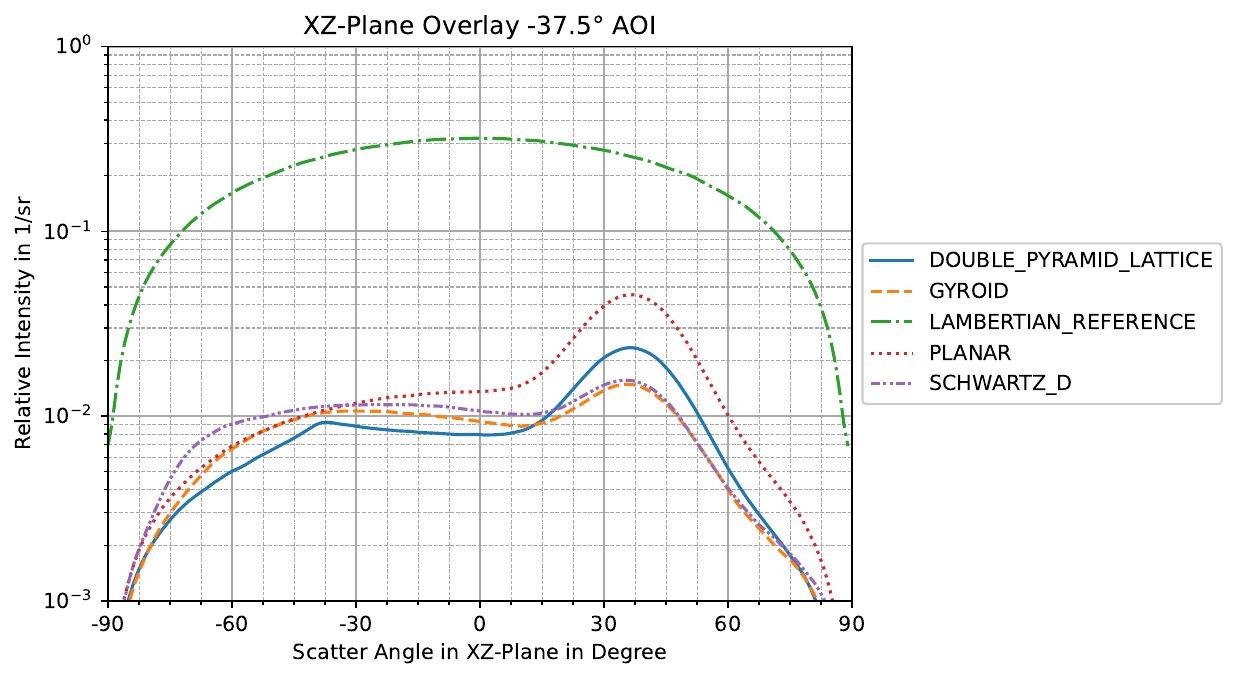}
    \end{minipage}
    \hfill
    \begin{minipage}[b]{0.49\textwidth}
        \centering
        \includegraphics[width=\linewidth]{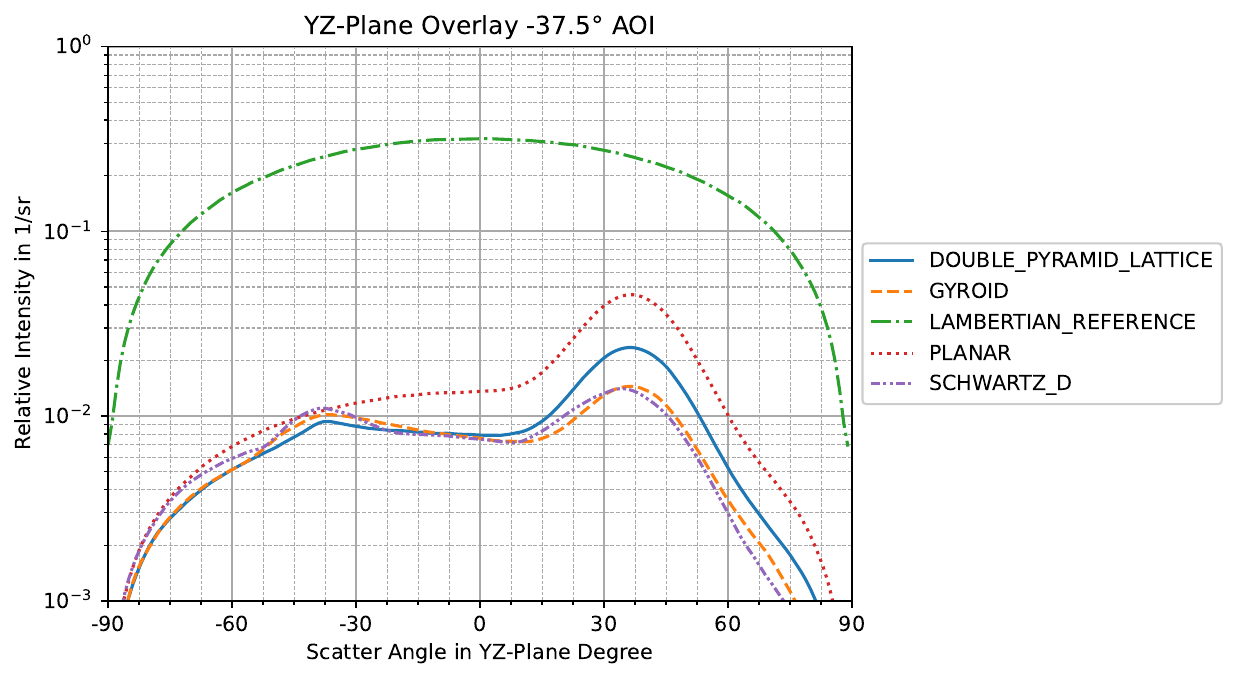}
    \end{minipage}
    
    \begin{minipage}[b]{0.49\textwidth}
        \centering
        \includegraphics[width=\linewidth]{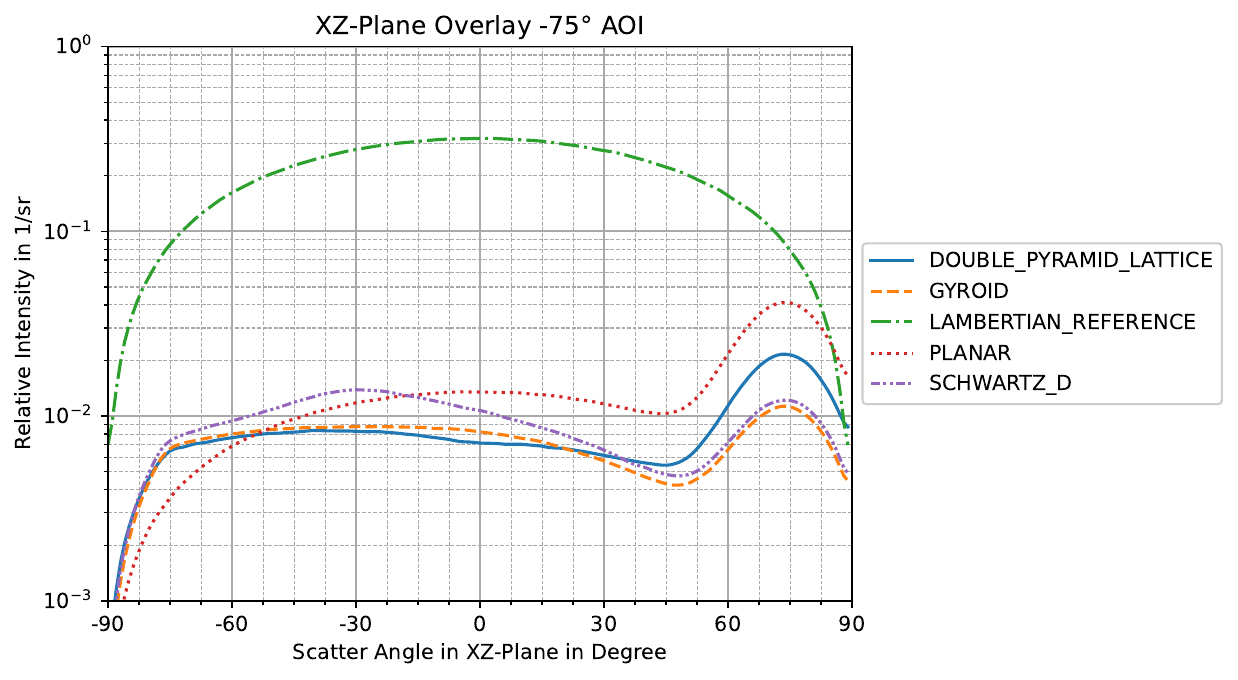}
    \end{minipage}
    \hfill
    \begin{minipage}[b]{0.49\textwidth}
        \centering
        \includegraphics[width=\linewidth]{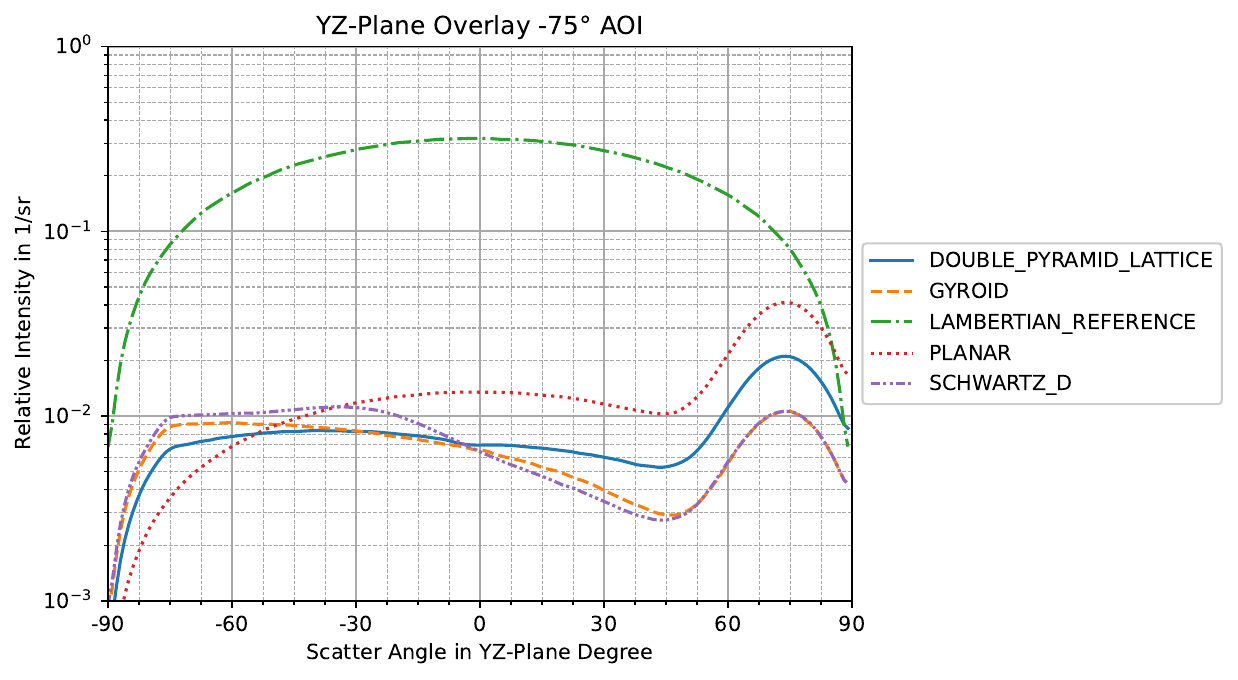}
    \end{minipage}

    \caption{Orthogonal sections in the XZ and YZ planes of the raw data results collected by the hemispherical far-field intensity receiver.}
    \label{fig:intesity_sections}
\end{figure}

\end{document}